\documentclass[aps,prd,floatfix,preprintnumbers,a4paper,nofootinbib,superscriptaddress,10pt,twocolumn]{revtex4-1}
\usepackage{lipsum}
\usepackage{multirow}
\usepackage[titletoc,title]{appendix}

\usepackage{float}
\usepackage[caption = false]{subfig}

\usepackage{graphicx}

\usepackage{amsmath}
\usepackage{amssymb}
\usepackage{amsfonts}
\usepackage{mathtools}
\usepackage{bm}
\usepackage[normalem]{ulem}

\usepackage[utf8]{inputenc}
\usepackage{newtxtext}

\usepackage[table]{xcolor}
\usepackage{url}
\usepackage[colorlinks, pdfborder={0 0 0}]{hyperref}
\definecolor{LinkColor}{rgb}{0.75, 0, 0}
\definecolor{CiteColor}{rgb}{0.75, 0, 0}
\definecolor{UrlColor}{rgb}{0, 0, 0.75}
\definecolor{bettergreen}{rgb}{0.0, 0.7, 0.1}
\hypersetup{linkcolor=LinkColor}
\hypersetup{citecolor=CiteColor}
\hypersetup{urlcolor=UrlColor}

\newcommand{\beq}{\begin{equation}}
\newcommand{\eeq}{\end{equation}}

\newcommand{\Cardiff}{School of Physics and Astronomy, Cardiff University, Queens Buildings, Cardiff, CF24 3AA, United Kingdom}
\newcommand{\Nikhef}{Nikhef, Science Park, 1098 XG Amsterdam, The Netherlands}
\newcommand{\AEIHannover}{Max Planck Institute for Gravitational Physics (Albert Einstein Institute), Callinstr. 38, 30167 Hannover, Germany}
\newcommand{\UniHannover}{Leibniz Universit\"at Hannover, D-30167 Hannover, Germany}
\newcommand{\Sapienza}{Dipartimento di Fisica, Universit\`{a} di Roma ``Sapienza'', Piazzale A. Moro 5, I-00185, Roma, Italy}
\newcommand{\INFN}{INFN Sezione di Roma, Piazzale A. Moro 5, I-00185, Roma, Italy}
\newcommand{\Potsdam}{Institut f\"{u}r Physik und Astronomie, Universit\"{a}t Potsdam, Haus 28, Karl-Liebknecht-Str. 24/25, 14476, Potsdam, Germany}

\usepackage[acronym,shortcuts]{glossaries}

\newacronym{bbh}{BBH}{black-hole binary}
\newacronym{bns}{BNS}{neutron-star binary}
\newacronym{nsbh}{NSBH}{neutron star-black hole}
\newacronym{eos}{EOS}{equation of state}
\newacronym{bh}{BH}{black hole}
\newacronym{EOB}{EOB}{effective-one body}

\newcommand{\NSBH}{\acs{nsbh} }
\newcommand{\BBH}{\acs{bbh} }
\newcommand{\EOS}{\acs{eos} }
\newcommand{\EOB}{\acs{EOB} }

\begin{document}

\title{Modeling the gravitational wave signature of\\ neutron star black hole coalescences: PhenomNSBH}

\author{Jonathan E. Thompson}
\affiliation{\Cardiff}
\author{Edward Fauchon-Jones}
\affiliation{\Cardiff}
\author{Sebastian~Khan}
\affiliation{\Cardiff}
\affiliation{\AEIHannover}
\affiliation{\UniHannover}
\author{Elisa~Nitoglia}
\affiliation{\Sapienza}
\affiliation{\INFN}
\author{Francesco Pannarale}
\affiliation{\Sapienza}
\affiliation{\INFN}
\author{Tim Dietrich}
\affiliation{\Nikhef}
\affiliation{\Potsdam}
\author{Mark Hannam}
\affiliation{\Cardiff}
\affiliation{\Sapienza}

\begin{abstract}
Accurate gravitational-wave (GW) signal models exist for \acf{bbh} and \acf{bns} systems, which are consistent with all of the published GW observations to date.
Detections of a third class of compact-binary systems, neutron-star-black-hole (NSBH) binaries, have not yet been confirmed, but are eagerly awaited in the near future. For NSBH systems, GW models
do not exist across the viable parameter space of signals. In this work we present the frequency-domain phenomenological model, \texttt{PhenomNSBH},
for GWs produced by \acs{nsbh} systems with mass ratios from equal-mass up to
15, spin on the \acf{bh} up to a dimensionless spin of \(|\chi|=0.5\), and tidal deformabilities ranging from 0 (the BBH limit) to 5000.
We extend previous work on a phenomenological amplitude
model for \acs{nsbh} systems to produce an amplitude model that is parameterized by a single tidal deformability parameter.
This amplitude model is combined with an analytic phase model describing tidal corrections. The resulting approximant is compared to publicly-available NSBH numerical-relativity simulations and hybrid waveforms constructed from numerical-relativity simulations and tidal inspiral approximants.
For most signals observed by second-generation ground-based detectors, it will be difficult to use the GW signal alone to distinguish single NSBH systems from either BNSs or BBHs, and therefore to unambiguously identify an NSBH system.
\end{abstract}

\date{\today}

\maketitle

%%%%%%%%%%%%%%%%%%%%%%%%%%%%%%%%%%%%%%%%%%%%%
\section{Introduction}
\label{sec:introduction}

Stellar-mass compact-binary coalescences have been the source of all current gravitational-wave (GW)
observations made by the Advanced LIGO~\cite{TheLIGOScientific:2014jea} and
Advanced Virgo detectors~\cite{TheVirgo:2014hva}. The data collected during the first and second observing runs is publicly available~\cite{Vallisneri:2014vxa,collaboration2019open}, and analyses of it have been published in several GW catalogues~\cite{LIGOScientific:2018mvr,Venumadhav:2019lyq,Nitz:2019hdf,Nitz:2019bxt}. The compact-binary mergers expected to be observed by current ground-based detectors come in three varieties: black-hole binaries (BBHs), neutron-star binaries (BNSs), and binaries that consist of one black hole and one neutron star (NSBHs). The majority of GW signals detected so far comes from BBH mergers, with two detections, GW170817~\cite{Abbott:2018wiz} and GW190425~\cite{Abbott:2020uma}, inferred to be from BNS mergers. Although the GW signals from these two events are also consistent with NSBH mergers, \textit{e.g.}, \cite{Coughlin:2019kqf, Kyutoku:2020xka}, this class of merger has yet to be unambiguously observed.

To extract physical information from GW signals, template waveforms constructed from theoretical models are compared with the data using a Bayesian framework.
Much of the previous waveform modeling efforts have focused successfully on BBHs --- for examples of recent BBH waveform models, see \texttt{SEOBNRv4HM}~\cite{Cotesta:2018fcv}, \texttt{PhenomPv3HM}~\cite{PhysRevD.100.024059,Khan:2019kot}, and surrogates \texttt{NRSur7dq4}~\cite{Varma:2019csw} and \texttt{NRHybSur3dq8}~\cite{PhysRevD.99.064045}. These BBH waveform models do not capture the changes to the waveform morphology introduced when one or both of the binary companions is a neutron star~(NS). One effect is a shift to the waveform phase that arises from tidal deformation of the NS during the inspiral of the two bodies~\cite{Flanagan:2007ix}.
This shift has been the focus of recent research into BNS waveform modeling efforts, and has produced several available models: \texttt{TEOBResumS}~\cite{Nagar:2018zoe}, \texttt{SEOBNRv4T}~\cite{PhysRevLett.116.181101,PhysRevD.94.104028,PhysRevD.100.024002}, and the \texttt{NRTidal} models~\cite{Dietrich:2017aum,Dietrich:2018uni,Dietrich:2019kaq}. These phase corrections have been sufficient in observations to date, because disruption of the NSs produces changes in the GW amplitude at high frequency~\cite{Clark:2015zxa,PhysRevD.100.044047,PhysRevD.100.104029}, where the detectors have been largely insensitive to the merger and post-merger BNS signal~\cite{Foucart:2018lhe,Dudi:2018jzn}.

In signals from NSBH systems, the phase shift during the inspiral stage due to NS tidal deformation is present, but it is unlikely that it will be observable with current detectors~\cite{Pannarale:2011pk}.  Further, and in contrast to BNS signals, merger and post-merger dynamics in NSBH systems are potentially accessible to current ground-based detectors due to these systems' potential for higher total masses, which can shift the GW signal at merger to a more sensitive part of the frequency band.  As the mass-ratio of the system increases, the merger morphology of the waveform can range from total disruption of the NS, in which case the amplitude of the waveform is exponentially suppressed at high frequency~\cite{PhysRevD.78.064054}, to non-disruptive signals for which the waveform is comparable to a BBH waveform, where the high-frequency amplitude is governed by the ringdown of the companion black hole (BH)~\cite{Foucart:2013psa}. Observations of the merger signal in an NSBH could allow us to place tighter constraints on the NS equation of state (EOS)~\cite{Kyutoku:2011vz, Kyutoku:2010zd, Pannarale:2015jia} and identify its source as an NSBH binary.
Of the waveform models existing currently, \texttt{LEA}~\cite{Lackey:2013axa} and the upgraded \texttt{LEA+} models are the only existing NSBH waveform models that include an NSBH-specific merger morphology and are calibrated against NSBH NR waveforms. While effective in their shared calibration range, their parameter space coverage is limited, in particular only to mass ratios between
2 and 5.

The aim of this work is to produce a new \acs{nsbh} model called \texttt{PhenomNSBH} that combines an
approximate reparameterization of the \acs{nsbh} amplitude model described by
\cite{Pannarale:2015jka} with the state-of-the-art tidal phase model described
in \cite{Dietrich:2019kaq}. As with previous work, the new model supports a spinning
BH with spin vector parallel to the orbital angular momentum of the
system and a non-spinning NS. Furthermore, we simplify the previous amplitude
modeling efforts by replacing dependence on the NS EOS with a single tidal
deformability parameter. This change is essential to allow our new model to be
used for parameter estimation. With these changes to the amplitude model and the integration of an improved phase description, our new model is valid over a larger parameter space and it is capable of generating accurate waveforms from equal mass up to mass-ratio 15.
At high mass ratios, the NS merges with the BH before disrupting, and the GW signal approaches that of
an equivalent BBH. As we show in Sec.~\ref{sec:analysis}, beyond mass-ratio 8 a BBH model will be sufficient for observations with a
signal-to-noise ratio less than 300.

In Sec.~\ref{sec:modelling} we describe and outline the waveform model \texttt{PhenomNSBH} presented in this paper, which is implemented as \texttt{IMRPhenomNSBH} in the open-source software package \texttt{LALSuite}~\cite{lalsuite}. To assess the \texttt{PhenomNSBH} model,
we compare it against numerical-relativity (NR) data for various NSBH systems in Sec.~\ref{sec:analysis}, presenting
alongside the same comparisons for other relevant waveform models, and we identify the regions of parameter space
where an NSBH model will be necessary to prevent measurement biases. Finally we
conclude with Sec.~\ref{sec:discussion}, where we summarize our results and
discuss directions for future work. In the remaining sections of this paper
geometric units are used such that $G=c=1$.

%%%%%%%%%%%%%%%%%%%%%%%%%%%%%%%%%%%%%%%%%%%%%
\section{Modelling \glsentrylong{nsbh} waveforms}
\label{sec:modelling}

In this section we present a model for the GW signal emitted by an \acs{nsbh} binary system that consists of a non-spinning NS and a BH with
spin angular momentum $\mathbf{S}_{\mathrm{BH}}$ parallel to the orbital angular momentum $\mathbf{L}$ of the system.
Such a system may be parameterized by four intrinsic parameters: $M$, the total mass of the system,
$M = M_{\mathrm{BH}} + M_{\mathrm{NS}}$, where $M_{\mathrm{BH}}$ and $M_{\mathrm{NS}}$ are the component masses
of the BH and NS, respectively; $q$, the mass ratio of the system where
$q = M_{\mathrm{BH}}/M_{\mathrm{NS}} \geq 1$; $\chi$, the dimensionless spin of the BH given by
$\chi = \mathbf{S}_{\mathrm{BH}}\cdot\hat{\mathbf{L}}/M_{\mathrm{BH}}^2$; and $\Lambda$,
the dimensionless NS tidal deformability parameter~\cite{Flanagan:2007ix,Hinderer:2007mb} defined in terms of the quadrupolar
Love number, \(k_2\), and compactness \(C = M_\text{NS}/R_\text{NS}\) of the NS,
\begin{align}
\label{eq:tidalD}
\Lambda & = \frac{2}{3}\frac{k_2}{C^5}.
\end{align}
We encapsulate these four parameters in the vector $\boldsymbol{\theta} = (M,q,\chi,\Lambda)$.
Note that, unlike \acs{bbh} models, the total mass $M$ cannot be separated as a scaling factor due to the
scale-dependent effects that arise in the waveform from the presence of the NS.

We seek a model of the complex strain in the frequency domain, $\tilde{h}(f;\boldsymbol{\theta},\vartheta,\varphi)$,
where the extrinsic parameters $(\vartheta, \varphi)$ represent the orientation of the system with respect to a distant observer.
The strain may be written as an expansion in spin-weighted spherical harmonics ${}_{-2}Y_{\ell m}(\vartheta, \varphi)$. For the first step in this preliminary model, we follow previous phenomenological models \cite{Santamaria:2010yb,Husa:2015iqa,Khan:2015jqa,Pannarale:2015jka}
and focus only on the dominant $(\ell,|m|) = (2,2)$ multipole moments, \textit{i.e.},
\begin{align}
\nonumber
\tilde{h}(f;\boldsymbol{\theta},\vartheta,\varphi) & = \sum_{\ell,m} \tilde{h}_{\ell m}(f;\boldsymbol{\theta}) \,{}_{-2}Y_{\ell m}(\vartheta, \varphi) \\
& \approx \sum_{m =\pm 2} \tilde{h}_{2m}(f;\boldsymbol{\theta})\, {}_{-2}Y_{2m}(\vartheta, \varphi).
\end{align}
The $\tilde{h}_{22}$ multipole moment is further decomposed in terms of an amplitude $A$ and phase $\phi$,
\begin{equation}
\label{eq:wfmodelAnsatz}
\tilde{h}_{22}(f;\boldsymbol{\theta})  = A(f;\boldsymbol{\theta})e^{-i \phi(f;\boldsymbol{\theta})},
\end{equation}
and we relate $\tilde{h}_{2-2}(f)=\tilde{h}_{22}^*(-f)$, where \(*\) denotes complex conjugation.
Higher multipoles are also necessary for unbiased parameter measurements
for systems with $q \geq 3$~\cite{Varma:2016dnf,Kalaghatgi:2019log}. A
quadrupole-only model is however sufficient to capture the broad phenomenology
of the signal from an NSBH system including the effects of tidal disruption, and
for all of the conclusions that we draw in this work. Using the calibrated higher-mode BBH model \texttt{IMRPhenomXHM}~\cite{Garcia-Quiros:2020qpx}, we estimate that the the first sub-dominant multipole moment contributes only an estimated 10-15\% of additional signal power when measured over the disruptive and mildly-disruptive region of the NSBH parameter space, due to the relatively low total mass of the NSBH system. We will discuss further
extensions in Sec.~\ref{sec:discussion}.

In the text that follows, we outline in detail how the amplitude and phase are modeled for an NSBH system.

%%%%%%%%%%%%%%%%%%%%%%%%%%%%%%%%%%%%%%%%%%%%%
\subsection{Amplitude model}
\label{subsec:amplitude}

To create an amplitude model for \texttt{PhenomNSBH} we start from the \NSBH amplitude description of
Pannarale \textit{et al.} in~\cite{Pannarale:2015jka}. This model describes an amplitude based on the aligned-spin \BBH
waveform amplitude of \texttt{PhenomC} \cite{Santamaria:2010yb}, which depends on three intrinsic parameters $(M, q,\chi)$
and an explicit choice of a NS \acf{eos}. Four choices of \EOS were used in its calibration, listed in order of increasing
softness, \textit{i.e.}, decreasing tidal deformability: 2H, H, HB, and B \cite{PhysRevD.79.124033}.
Given an \EOS and NS gravitational mass \(M_\text{NS}\) (assuming \(M_\text{NS}\le M_\text{BH}\)),
the amplitude model of Pannarale \textit{et al.} integrates the Tolman-Oppenheimer-Volkoff
equations~\cite{Tolman169,PhysRev.55.364,PhysRev.55.374} to find the NS radius \(R_\text{NS}\) and baryonic mass \(M_\text{b,NS}\) associated
with its gravitational mass. From the mass and radius, the NS compactness is computed via
\(C=M_\text{NS}/R_\text{NS}\).

While determination of the NS \EOS may be possible after several detections \cite{Lackey:2014fwa},
it is more practical for our waveform model to not be directly dependent on the \EOS. To this end, we replace the dependency of the amplitude model
on the EOS with a dependency on the dimensionless tidal deformability \(\Lambda\), outlined in Appendix~\ref{app:EOS}. With these augmentations made to the original amplitude model, we have a working amplitude for an aligned-spin \NSBH
system with dependence on the four intrinsic parameters \((M, q, \chi, \Lambda)\). Based on the workflows provided in
Ref.~\cite{Pannarale:2013uoa,Pannarale:2015jka}, the amplitude model is evaluated using the following steps:

\begin{table*}[htbp]
\setlength\tabcolsep{0.01\textwidth}
\renewcommand{\arraystretch}{2.0}
\begin{tabular}{r | p{0.165\textwidth} | p{0.165\textwidth} | p{0.33\textwidth} | p{0.18\textwidth}}
\hline\hline \rule{0pt}{5ex}
\multirow{3}{*}{\rotatebox{90}{\textbf{Merger type}\;\;}} & \multicolumn{1}{c}{\parbox{0.165\textwidth}{\textbf{non-disruptive} \\ \emph{(no torus remnant)}}}
  & \multicolumn{1}{c}{\parbox{0.165\textwidth}{\textbf{mildly disruptive} \\ \emph{(torus remnant)}}}
  & \multicolumn{1}{c}{\parbox{0.33\textwidth}{\textbf{mildly disruptive} \\ \emph{(no torus remnant)}}}
  & \multicolumn{1}{c}{\parbox{0.18\textwidth}{\textbf{disruptive} \\ \emph{(torus remnant)}}} \\ [2.0ex]
\cline{2-5}
& \multicolumn{2}{c|}{$f_{\mathrm{tide}} \geq f_{\mathrm{RD}}$}
  & \multicolumn{2}{c}{$f_{\mathrm{tide}} < f_{\mathrm{RD}}$} \\
\cline{2-5}
& \multicolumn{1}{c|}{$M_{\mathrm{b,torus}} = 0$}
  & \multicolumn{1}{c|}{$M_{\mathrm{b,torus}} > 0$}
  & \multicolumn{1}{c|}{$M_{\mathrm{b,torus}} = 0$}
  & \multicolumn{1}{c}{$M_{\mathrm{b,torus}} > 0$} \\
\hline\hline
$\epsilon_{\mathrm{tide}}$ &  \multicolumn{2}{c|}{$\omega^{+}_{x_1,d_1}(x_{\mathrm{ND}}) \; [x_1\!=\!-0.0796251, d_1\!=\!0.0801192]$} & 0.0 & 0.0  \\ \cline{2-5}
$\epsilon_{\mathrm{ins}}$ & 1.0 & \multicolumn{3}{c}{$1.29971 - 1.61724x_D$} \\ \cline{2-5}
$\sigma_{\mathrm{tide}}$ & \multicolumn{2}{c|}{$\omega^{-}_{x_2,d_2}(x'_{\mathrm{ND}}) \; [x_2\!=\!-0.206465 , d_2\!=\!0.226844]$}& $(\omega^{-}_{x_2,d_2}(x'_{\mathrm{ND}}) + 0.137722 - 0.293237x'_{\mathrm{D}})/2$ & $0.137722 - 0.293237x'_{\mathrm{D}}$  \\ \cline{2-5}
$\tilde{f}_0$ & $\tilde{f}_{\mathrm{RD}}$ & $\epsilon_{\mathrm{ins}} \tilde{f}_{\mathrm{RD}}$ & $[(q - 1)\tilde{f}_{\mathrm{RD}} + \epsilon_{\mathrm{ins}} f_{\mathrm{tide}}]/q$ & $\epsilon_{\mathrm{ins}} f_{\mathrm{tide}}$  \\ \cline{2-5}
$\tilde{f}_1$ & $\tilde{f}_{\mathrm{RD}}$ & $\epsilon_{\mathrm{ins}} \tilde{f}_{\mathrm{RD}}$ & $[(q - 1)\tilde{f}_{\mathrm{RD}} + f_{\mathrm{tide}}]/q$ & $f_{\mathrm{tide}}$ \\ \cline{2-5}
$\tilde{f}_2$ & \multicolumn{2}{c|}{$\tilde{f}_{\mathrm{RD}}$} & $-$ & $-$ \\ [0.5ex]
\hline\hline
\end{tabular}
\caption{\label{tab:merger-dep-quants} Summary of merger type dependent components of the amplitude model. For the definitions of $x_{\mathrm{ND}}$,  $x'_{\mathrm{ND}}$,  $x_{\mathrm{D}}$ and $x'_{\mathrm{D}}$, see Eqs.~(\ref{eq:x_ND})-(\ref{eq:x_D_prime}) in Appendix~\ref{app:amplitude}. Note that all applications of window functions \(\omega^\pm\) for merger-type dependent quantities are a factor of two smaller to correct for a typographical error in \cite{Pannarale:2015jka}. The adjusted ringdown frequency is defined as \(\tilde{f}_{\mathrm{RD}} = 0.99 \times 0.98 f_{\mathrm{RD}}\) for $\Lambda > 1$ and \(\tilde{f}_{\mathrm{RD}} = 0.98 f_{\mathrm{RD}}\) for $\Lambda = 0$ with a smooth interpolation given by Eq.~(\ref{eq:f_RD_tilde}).}
\end{table*}

\begin{enumerate}

\item \textbf{Calculate the NS compactness $C$} \\
Evaluate Eq.~(\ref{eq:compactness}) to calculate compactness $C(\Lambda)$ of the
NS.

\item \textbf{Calculate the tidal disruption frequency $f_{\mathrm{tide}}$} \\
Evaluate Eq.~(\ref{eq:ftide}) to calculate the tidal disruption frequency
$f_{\mathrm{tide}}(q, \chi, C)$.

\item \textbf{Calculate the baryonic mass ratio $M_{\mathrm{b,torus}}/M_{\mathrm{b,NS}}$} \\
Evaluate Eq.~(\ref{eq:baryonic-mass-ratio}) to calculate the baryonic mass ratio
of the NS. This model depends only on the torus remnant baryonic mass
$M_{\mathrm{b,torus}}$ and the baryonic mass $M_{\mathrm{b,NS}}$ of the isolated
NS at rest through expressions of the form
$M_{\mathrm{b,torus}}/M_{\mathrm{b,NS}}$. As such it is not necessary to
calculate an explicit value for $M_{\mathrm{b,NS}}$, which was required by
\cite{Pannarale:2015jka}.

\item \textbf{Calculate remnant BH properties $(\chi_f, M_f)$} \\
Evaluate Eq.~(\ref{eq:remnant-model}) to calculate the final spin $\chi_f(\eta,
\chi, \Lambda)$ and final mass $M_f(\eta, \chi, \Lambda)$ of the remnant black
hole, where $\eta = q/(1+q)^2$ is the symmetric mass ratio.

\item \textbf{Calculate the remnant BH quantities ($f_{\mathrm{RD}}, Q$)} \\
Evaluate Eqs.~(\ref{eq:ringdown-frequency}) and (\ref{eq:quality-factor}) to
calculate the ringdown frequency $f_{\mathrm{RD}}(M_f, \chi_f)$ and quality
factor $Q(\chi_f)$.

\item \textbf{Calculate merger-type dependent quantities} \\
Calculate the merger-type dependent quantities $(\epsilon_{\mathrm{tide}},
\epsilon_{\mathrm{ins}}, \sigma_{\mathrm{tide}}, \tilde{f}_0, \tilde{f}_1,
\tilde{f}_2)$ using the conditions on $f_{\mathrm{tide}}$, $f_{\mathrm{RD}}$,
and the magnitude of $M_{\mathrm{b,torus}}$, and expressions provided in
Table~\ref{tab:merger-dep-quants}.

\item \textbf{Calculate non-merger-type dependent quantities} \\
Evaluate Eq.~(\ref{eq:phenomc-param-model}) to calculate the phenomenological
parameters $\gamma_1$, $\delta_1$, and $\delta_2$. Evaluate Eqs.~(\ref{eq:gamma-1-prime}) and (\ref{eq:delta-2-prime}) to calculate the phenomenological correction parameters $\gamma'_1$ and $\delta'_2$, respectively. While $\delta_1$ and $\delta'_2$ are not explicitly dependent on any merger-type dependent quantities, they are not required if the onset of tidal disruption happens before the ringdown frequency is reached.

\item \textbf{Evaluate the amplitude} \\
Evaluate the amplitude $A(f;\boldsymbol{\theta})$,
\begin{align}
\nonumber
A(f) & = A_{\mathrm{PN}}(f) \omega_{\tilde{f}_0,0.015+\sigma_{\mathrm{tide}}}^{-}(f) \\
\nonumber
& \quad+ \gamma'_1f^{5/6} \omega_{\tilde{f}_1,0.015+\sigma_{\mathrm{tide}}}^{-}(f) \\
& \quad+ A_{\mathrm{RD}}(f) \omega_{\tilde{f}_2,0.015+\sigma_{\mathrm{tide}}}^{+}(f),
\label{eq:ampansatz}
\end{align}
where $\omega_{f_0,d}^{\pm}$ is defined by Eq.~(\ref{eq:omega_window}) and
$A_{\mathrm{PN}}$ and $A_{\mathrm{RD}}$ are defined by
Eqs.~(\ref{eq:A_PN}) and (\ref{eq:A_RD}), respectively. We have suppressed all explicit
parametrization in the component functions of the amplitude \(A\) for
legibility.
\end{enumerate}

The specific parameters used in the amplitude model change based on the classification of the system, as detailed in Table~\ref{tab:merger-dep-quants}. As described in Refs.~\cite{Pannarale:2013uoa,Pannarale:2015jka}, the type of merger modeled by the amplitude varies between non-disruptive and disruptive cases, with the conditions identifying each case listed at the top of Table~\ref{tab:merger-dep-quants}. We now briefly summarize the various cases allowed for in the model. When the computed tidal disruption frequency, \(f_\text{tide}\), exceeds the ringdown frequency, \(f_\text{RD}\), and the model predicts no remnant torus mass, \(M_\text{b,torus}=0\), the signal is classified to be \emph{non-disruptive}, as it is assumed that the binary merges before the NS is tidally disturbed to the point of disruption. If \(f_\text{tide}<f_\text{RD}\), the effects of disruption appear earlier in the waveform and the system transitions toward disruption. In this case, the torus mass remnant is used to distinguish between systems that disrupt early and form a remnant mass disk, called \emph{disruptive} systems, and those that disrupt late in the inspiral and no disk forms, labeled \emph{mildly disruptive}. The fourth case in the table, when \(f_\text{tide}\ge f_\text{RD}\) and the system predicts a non-zero torus mass remnant, is a result of shortcomings in the fitting formulae~\cite{Pannarale:2015jka}, and has not arisen in our dense sampling of waveforms across the broad parameter space listed in this paper.

The torus mass and ringdown conditions that partition the parameter space into the different merger types are non-linear in the intrinsic parameters of \texttt{PhenomNSBH} and the boundaries between the different regions of the model cannot be explicitly written. Fig. 2 provides an example of the transition between these cases as the mass-ratio of the system increases for fixed tidal deformability and NS mass. For a full discussion on the distribution of merger types across the parameters please see~\cite{Pannarale:2015jka}. A more detailed description of the amplitude model workflow is given in
Appendix~\ref{app:amplitude}. For full details of the amplitude model, along
with the different merger types, we direct the reader to Sec.~IV of
Ref.~\cite{Pannarale:2015jka}.

%%%%%%%%%%%%%%%%%%%%%%%%%%%%%%%%%%%%%%%%%%%%%
\subsection{Phase model}

In addition to a proper amplitude description, we need to model the GW phase \(\phi\) for the \NSBH coalescence in such a way that it provides an accurate description within a large region of the parameter space and incorporates tidal effects imprinted in the signal.
As a \BBH baseline, we use the frequency-domain phase approximant from \texttt{PhenomD}\ \cite{Husa:2015iqa,Khan:2015jqa}.
This model allows for a description of \BBH systems up to mass ratios of \(q\le18\) and aligned-spin components up to \(|\chi|\le0.8\). We augment this BH baseline with tidal effects modeled within the \texttt{NRTidal} approach~\cite{Dietrich:2017aum,Dietrich:2018uni},
using the newest version as described in Ref.~\cite{,Dietrich:2019kaq}.
The \texttt{NRTidal} phase model includes matter effects in the form of a closed-form, analytical expression,
combining post-Newtonian knowledge with \EOB and NR information. While this model was designed to be an accurate phase model for BNS systems, recent work \cite{Foucart:2018lhe} has shown that it is also a valid description in the \NSBH limit.

\begin{table}[h]
\begin{ruledtabular}
\begin{tabular}{l l l l l l l l}
Name & SXS Name &  \(q\) & \(M_\text{BH}\) & \(M_\text{NS}\) & \(\chi_\text{NS}\)  & \(\Lambda\)& Merger Type\\
\hline
q1a0 &SXS:BHNS:0004&1 & 1.4 &1.4& 0 &791&Disruptive\\
q1.5a0 &SXS:BHNS:0006&1.5 & 2.1 &1.4& 0 &791&Disruptive\\
q2a0 &SXS:BHNS:0002& 2 & 2.8 &1.4& 0 &791&Disruptive\\
q3a0 &SXS:BHNS:0003&3 & 4.05 &1.35& 0 &607&Mildly Disruptive\\
q6a0 &SXS:BHNS:0001& 6 & 8.4 &1.4& 0 &525&Non-disruptive\\
q1a2 &SXS:BHNS:0005&1 & 1.4 &1.4& -0.2 &791&Disruptive\\
q2a2 &SXS:BHNS:0007&2 & 2.8 &1.4& -0.2 &791&Disruptive\\
\end{tabular}
\caption{SXS waveforms \cite{Foucart:2013psa, Chakravarti:2018uyi, Foucart:2018lhe} and their parameters used for comparisons and in making the hybrids. Along with the name given in the SXS public catalog, we also list an abbreviated name given to each waveform in this paper.
\label{T:SXSWaveformTable}}
\end{ruledtabular}
\end{table}

%%%%%%%%%%%%%%%%%%%%%%%%%%%%%%%%%%%%%%%%%%%%%
\section{Analysis of model}
\label{sec:analysis}

To quantify the effectiveness of our model at reproducing \NSBH waveforms, we compare against a selection of NR \NSBH waveforms produced by the SXS collaboration \cite{Foucart:2013psa, Chakravarti:2018uyi, Foucart:2018lhe} with simulation parameters listed in Table~\ref{T:SXSWaveformTable}. To carry out these comparisons, it is useful to introduce the notion of the \textit{overlap} between two waveforms \(h_1\) and \(h_2\),
\begin{equation}
\left<h_1|h_2\right>=4\Re\int_{f_1}^{f_2}\frac{\tilde{h}_1(f)\tilde{h}^*_2(f)}{S_n(f)}\text{d}f,
\label{overlap}
\end{equation}
which is the functional inner-product weighted by the detector noise power-spectral density, \(S_n(f)\), taken
for this work to be the Advanced LIGO zero-detuned, high-power (AZDHP) noise curve~\cite{aligozerodethp}, which is the current
goal for the detector's design sensitivity. By maximizing the normalized overlap over phase ($\phi_c$) and time ($t_c$) shifts to \(h_1\), one
determines the \textit{faithfulness} with which \(h_1\) represents \(h_2\),
\begin{equation}
\mathcal{F}=\max_{\phi_c,t_c}\frac{\left<h_1(\phi_c,t_c)|h_2\right>}{||h_1|| ||h_2||},
\label{faithfulness}
\end{equation}
where \(||h||^2=\left<h|h\right>\).

In the following subsection we compare PhenomNSBH with publicly available numerical relativity waveforms for systems with non-spinning black holes. As there are no publicly available NR waveforms with non-zero black hole spin, such as were included in the set of simulations used in the calibration of both the \texttt{LEA+} model~\cite{Lackey:2013axa} and the amplitude model~\cite{Pannarale:2015jka}, we first perform a comparison with LEA+ to analyse the faithfulness of the model including when the black hole has spin. We note that there is a small region above $q=5$ for high BH spin and high tidal deformability where disruption may occur. This represents a region of extrapolation for the amplitude fits; while we believe that this extrapolation is well-behaved and reasonable, there are currently no NR simulations against which we can test the model in this small region of parameter space.

The original \texttt{LEA} model was constructed as a phenomenological NSBH model from baseline \texttt{PhenomC}~\cite{Santamaria:2010yb} and \texttt{SEOBNR}~\cite{PhysRevD.86.024011} BBH waveform models. Additions to the BBH models were made to include tidal PN terms during the inspiral, and a taper was applied to the merger contributions of the waveform that was calibrated against NSBH NR waveforms. The \texttt{LEA+} model was introduced as an improvement to the \texttt{LEA} model by substituting a reduced-order model of \texttt{SEOBNRv2}~\cite{PhysRevD.89.061502} for the underlying BBH waveform. The \texttt{LEA+} model is calibrated for NS masses ranging between \(1.2-1.4M_\odot\), mass-ratios \(q\in[2,5]\), and BH spins \(-0.5\le\chi\le0.75\). To perform the comparison, we generate  waveforms across the overlapping parameter spaces covered by the calibration ranges of \texttt{LEA+} and \texttt{PhenomNSBH} and compute the faithfulness between waveforms generated using identical parameters.  The results show good agreement between the models, with $\mathcal{F}>0.99$. The comparison only deviates noticeably when \(\chi<-0.4\), where the faithfulness drops to 0.98.

Finally, we remark here that the testing done against numerical relativity is 
performed with simulations utilizing tidal deformabilities below 1000. 
While the amplitude and phase models were individually calibrated with 
simulations where \(\Lambda\) extends to above 4000, which motivates 
the coverage of \(\Lambda\) that we provide for this model, this 
calibration of the amplitude was limited in mass-ratio to between 
2-5 and for NS with a limited range of masses. The good agreement 
with simulation q1a0 included in the NR comparisons below 
with a tidal deformability of 791 demonstrates the ability for the 
model to extrapolate to equal-mass configurations, however we may expect 
physical effects to dominate at \textit{e.g.}, high tidal deformability 
and either equal mass or high NS mass, which have not been captured by this 
calibration, and we would recommend caution when using the model in this 
regime.

%%%%%%%%%%%%%%%%%%%%%%%%%%%%%%%%%%%%%%%%%%%%%
\subsection{Comparison to numerical relativity}

NR simulations typically cover the last orbits before coalescence.
For the \NSBH NR waveforms we consider in validating the model, the typical starting GW
frequency is between 300--400~Hz and covers between 10 and 16 orbits before
merger.
Currently Advanced LIGO and Virgo are sensitive to signals starting around
$20 \, \rm{Hz}$, which for a true signal will include on the order of \(10^3\) orbits prior to merger,
 and therefore the NR waveforms used here are
missing a large portion of the inspiral signal~\cite{Ohme:2011zm}.
We will address this issue by constructing hybrid waveforms for comparison against the model;
the results of a comparison against hybrid
waveforms can be found in Sec.~\ref{sec:hybrids}.
We first compare against the NR data directly in order to assess the accuracy
of the model during the late-inspiral and merger.

The results from comparing directly
with the NR waveforms are given in Table~\ref{T:NRMatches},
and the faithfulness is computed over the frequency range covered
by each NR waveform. We provide results from using the
AZDHP (design) noise curve,
as well as a flat noise curve (in parentheses).
We also compute the faithfulness of several other
waveform models to gauge the systematic uncertainty that
is incurred by using them.
Specifically, we also compare against the \NSBH model \texttt{LEA+}~\cite{Lackey:2013axa},
an inspiral \NSBH model
\texttt{SEOBNRv4T}~\cite{PhysRevLett.116.181101,PhysRevD.94.104028},
a BBH model
\texttt{PhenomD}~\cite{Husa:2015iqa,Khan:2015jqa}
and two inspiral BNS models
\texttt{PhenomDNRT}~\cite{Dietrich:2018uni,Dietrich:2019kaq,Husa:2015iqa,Khan:2015jqa} and
\texttt{SEOBNRv4NRT}~\cite{Dietrich:2018uni,Dietrich:2019kaq,2017PhRvD..95d4028B}~\footnote{The approximant names in the \texttt{LALSuite} code
for \texttt{LEA+}, \texttt{PhenomD}, \texttt{PhenomDNRT}, \texttt{SEOBNRv4T} and
\texttt{SEOBNRv4NRT} are
\texttt{Lackey\_Tidal\_2013\_SEOBNRv2\_ROM}, \texttt{IMRPhenomD}, \texttt{IMRPhenomD\_NRTidalv2},
\texttt{SEOBNRv4T} and \texttt{SEOBNRv4\_ROM\_NRTidalv2}, respectively.}.

\begin{figure*}[ht!]
\subfloat{\includegraphics[width=0.5\textwidth]{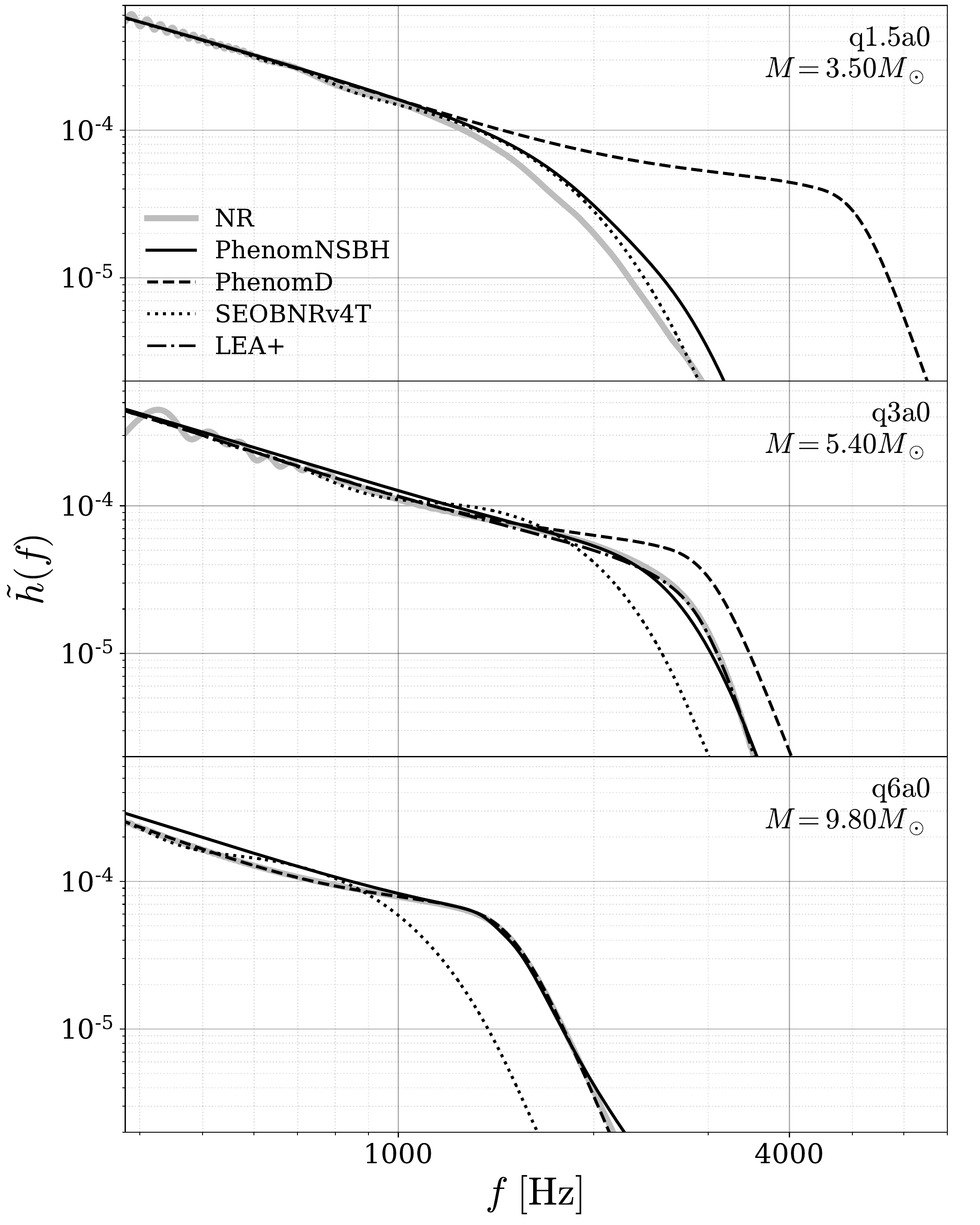}}
\subfloat{\includegraphics[width=0.5\textwidth]{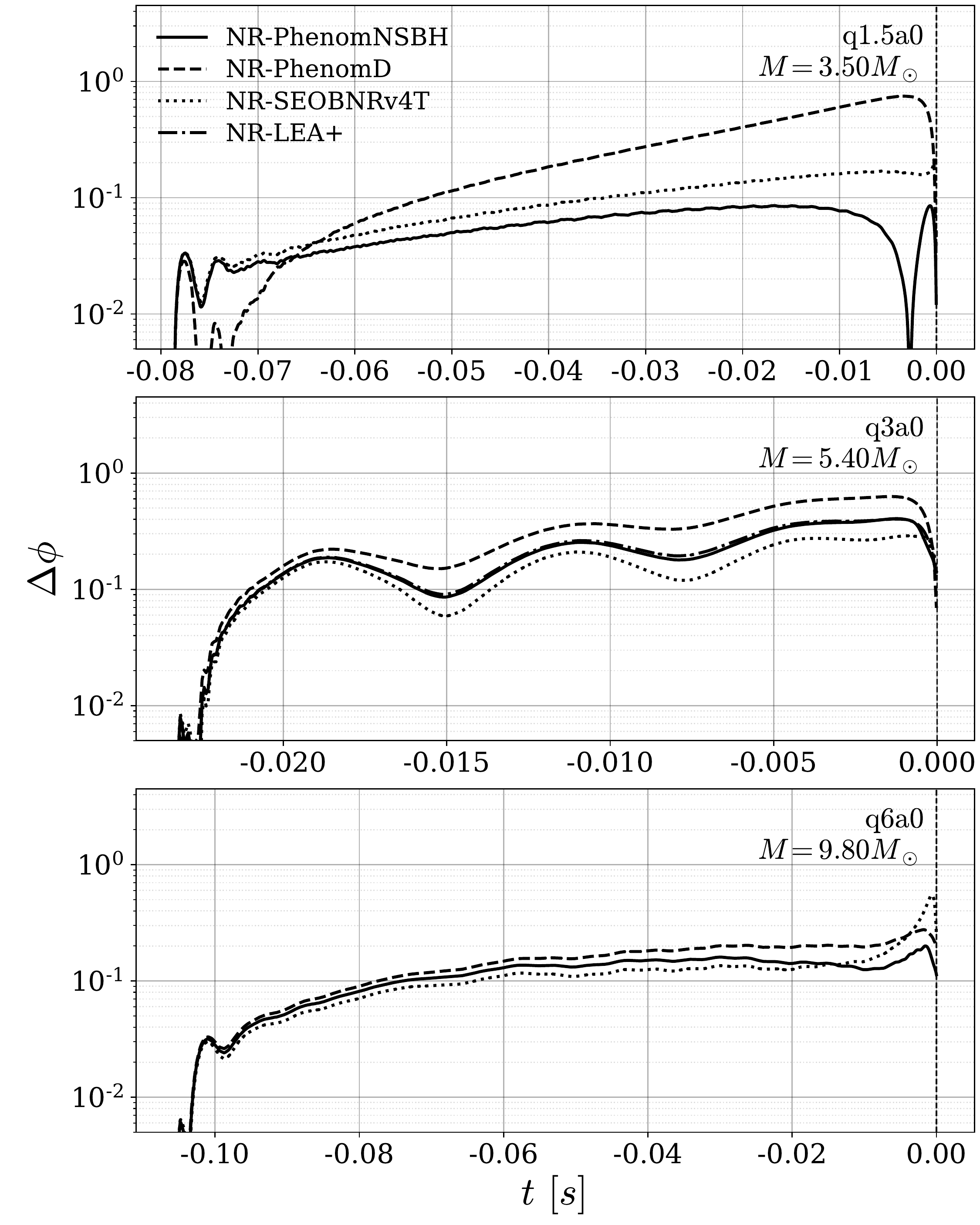}}
\caption{Comparisons between NR waveforms and various models. The left-most plots show \(\tilde{h}(f)\) for each NR case over the last few orbits before merger, along with frequency-domain representations of the signal using various approximants. The right-most plots display the accumulated time-domain phase error between the NR phase and each approximant over the length of the NR data, and the approximant signals are aligned by time- and phase-shifts to the NR data over a few GW cycles near the start of each NR simulation. Only the \texttt{q3a0} case falls within the parameter space coverage of the \texttt{LEA+} model.\label{fig:TD-comp}}
\end{figure*}

In Ref.~\cite{Foucart:2018lhe} the authors analyse the same NR
waveforms and the same models. We find similar results
and plot these in Fig.~\ref{fig:TD-comp}. Although that work focuses on
the agreement between the model and NR by studying the de-phasing,
here we focus on computing the faithfulness, which is directly
related to the loss in signal-to-noise ratio in matched-filter
based searches, and takes into account both phase and amplitude differences.

\begin{table*}[ht]
\begin{ruledtabular}
\begin{tabular}{lllllll}
\hline
 Sim Name & PhenomNSBH & PhenomD    & PhenomDNRT    & SEOBNRv4NRT   & SEOBNRv4T     & LEA+          \\
 q1a0     & 0.988 (0.978) & 0.911 (0.834) & 0.986 (0.972) & 0.988 (0.976) & 0.997 (0.994) & -             \\
 q1.5a0   & 0.997 (0.994) & 0.955 (0.906) & 0.998 (0.995) & 0.998 (0.995) & 0.999 (0.997) & -             \\
 q2a0     & 0.999 (0.997) & 0.973 (0.931) & 0.994 (0.983) & 0.994 (0.983) & 0.997 (0.994) & 0.999 (0.997) \\
 q3a0     & 0.994 (0.990) & 0.984 (0.971) & 0.929 (0.841) & 0.930 (0.842) & 0.983 (0.963) & 0.994 (0.994) \\
 q6a0     & 0.999 (0.998) & 0.999 (0.999) & 0.893 (0.842) & 0.893 (0.842) & 0.983 (0.966) & -             \\
 q1a2     & 0.894 (0.844) & 0.809 (0.701) & 0.885 (0.822) & 0.888 (0.826) & 0.900 (0.850) & -             \\
 q2a2     & 0.986 (0.974) & 0.947 (0.900) & 0.992 (0.985) & 0.994 (0.988) & 0.985 (0.969) & - \\
\hline
\end{tabular}
\caption{The computed faithfulness between the seven SXS NSBH numerical relativity
waveforms and the waveform approximants
\texttt{PhenomNSBH}, \texttt{PhenomD}, \texttt{PhenomDNRT}, \texttt{SEOBNRv4T}, \texttt{SEOBNRv4NRT}, and \texttt{LEA+}.
We compute two sets of matches. The first uses the Advanced LIGO zero-detuning, high-power noise curve
and second, in parentheses, uses a flat noise curve. The frequency range used
to compute the matches cover the entire bandwidth of the NR data.
\label{T:NRMatches}
}
\end{ruledtabular}
\end{table*}

Of the seven NR simulations considered in this paper, five are binary systems without any spin on either body (see Table~\ref{T:SXSWaveformTable} for a list of the non-spinning waveforms and their parameters). The two cases including spin, q1a2 and q2a2, are simulations
where the NS is spinning with a dimensionless spin magnitude of
$0.2$ in a direction anti-parallel to the orbital angular momentum. The amplitude model used for \texttt{PhenomNSBH} is not calibrated for spinning NSs, however it is constructed to propagate the NS spin through to the uderlying BNS tidal phase model~\cite{Dietrich:2019kaq} and the underlying BBH amplitude model~\cite{Santamaria:2010yb} where the NS spin is treated as a BH spin. These two NR waveforms with spinning NS allow for an exploration of the viability of the model when the NS is spinning. We do not make direct comparisons to NR where the BH is spinning as no such simulations are currently publicly available. The amplitude model used in this work was calibrated against NSBH NR waveforms with a spinning BH. Furthermore, based on the faithfulness comparisons with \texttt{LEA+}, which is also calibrated to and validated against the same NSBH NR waveforms with a spinning BH, we expect the model to also perform well where the BH is spinning and provide an accurate model for these systems up to numerical errors present in the original calibration set for these models.

For q1a2, when we compare against the
BBH model \texttt{PhenomD}
the match is 0.809 (0.701) for the AZDHP (flat) noise curve.
Including tidal effects in the model does improve the match
where we find a match
of $\sim 0.89$ ($\sim 0.84$) for the AZDHP (flat) noise curve.
For q2a2, the match is not as bad as q1a2
but the results are, in general, worse
than the non-spinning cases. Comparisons against \texttt{LEA+} are not included for these two NSBH NR waveforms with a spinning NS as \texttt{LEA+} does not depend on NS spin.

Reference~\cite{Foucart:2018lhe} showed that the NR phase error is smaller than
the systematic modelling error in the original \texttt{NRTidal} phase
approximant model. Similarly, we also find a noticeable phase difference between
the phase description employed in \texttt{PhenomNSBH} and the NR data. These
results suggest that further improvements such as a new phase calibration to
NSBH NR simulations or the inclusion of spin-dependent f-mode resonance shifts
near merger~\cite{PhysRevLett.116.181101} may be important to include.
In the next Section, however, we show that the measured dephasing is not an
issue for Advanced LIGO at design sensitivity.

%%%%%%%%%%%%%%%%%%%%%%%%%%%%%%%%%%%%%%%%%%%%%
\subsection{Comparison to hybrid numerical-relativity waveforms}
\label{sec:hybrids}

\begin{table*}[ht]
\begin{ruledtabular}
\begin{tabular}{lllllll}
\hline
 Sim Name & PhenomNSBH   & PhenomD      & PhenomDNRT      & SEOBNRv4NRT     & SEOBNRv4T       & LEA+            \\
 q1a0     & 0.9996 (0.9996) & 0.9906 (0.9936) & 0.9985 (0.9989) & 0.9992 (0.9994) & 0.9968 (0.9982) & -               \\
 q1.5a0   & 0.9994 (0.9997) & 0.9930 (0.9952) & 0.9991 (0.9993) & 0.9979 (0.9984) & 0.9973 (0.9981) & -               \\
 q2a0     & 0.9987 (0.9990) & 0.9954 (0.9966) & 0.9989 (0.9993) & 0.9969 (0.9978) & 0.9970 (0.9976) & 0.9997 (0.9998) \\
 q3a0     & 0.9995 (0.9997) & 0.9956 (0.9975) & 0.9990 (0.9993) & 0.9975 (0.9984) & 0.9993 (0.9995) & 0.9990 (0.9990) \\
 q6a0     & 0.9974 (0.9981) & 0.9964 (0.9972) & 0.9946 (0.9974) & 0.9957 (0.9972) & 0.9977 (0.9988) & -               \\
 q1a2     & 0.9969 (0.9978) & 0.9405 (0.9508) & 0.9949 (0.9967) & 0.9962 (0.9972) & 0.9965 (0.9975) & -               \\
 q2a2     & 0.9991 (0.9992) & 0.9806 (0.9837) & 0.9985 (0.9992) & 0.9988 (0.9990) & 0.9982 (0.9989) & - \\
\hline
\end{tabular}
\caption{The computed faithfulness between the seven SXS NSBH numerical relativity
hybrid waveforms.
These have been hybridized with the \texttt{TEOBResumS} model with a start frequency of $20 \, \rm{Hz}$.
We compare against
the waveform approximants
\texttt{PhenomNSBH}, \texttt{PhenomD}, \texttt{PhenomDNRT}, \texttt{SEOBNRv4T}, \texttt{SEOBNRv4NRT}, and \texttt{LEA+}.
We compute two sets of matches. The first uses the Advanced LIGO zero-detuning, high-power noise curve
and second, in parentheses, uses a flat noise curve. The frequency range used
to compute the matches cover the entire bandwidth of the hybrid waveforms, down to a lower frequency bound of 20~Hz.
\label{T:HybridMatches}
}
\end{ruledtabular}
\end{table*}

We now repeat the comparisons performed above, but we use hybridized NR waveforms
to test the accuracy of the models for realistic signals including
the thousands of inspiral cycles prior to merger.
To do this, we produce hybrid waveforms, attaching the SXS \NSBH waveforms listed in
Table~\ref{T:SXSWaveformTable} to the tidal inspiral
approximant \texttt{TEOBResumS} \cite{Nagar:2018zoe},
following the hybridization procedure outlined in
~\cite{Hotokezaka:2016bzh, Dietrich:2018uni}. These hybrids have a starting frequency below $20 \, \rm{Hz}$
and allow us to test the models in a realistic observational scenario
where a current-generation ground-based detector would also be sensitive to the full inspiral from $20 \, \rm{Hz}$;
for the faithfulness integrals
we use a low frequency cutoff of $20 \, \rm{Hz}$. We have verified
the accuracy of our hybrid construction method and find that the
mismatch of a given hybrid with respect to itself subject
to varying the hybridization parameters is \(\mathcal{O}(10^{-4})\).

We list the results of the faithfulness calculations in Table~\ref{T:HybridMatches}. In general we find that the matches are very high,
even when comparing the NSBH hybrids against BBH models, with the exception of the spinning NSBH waveform \texttt{q1a2}.
At the total masses considered here, the signal-to-noise ratio (SNR) detectable
in Advanced LIGO is dominated by the long inspiral, and as a result
inaccuracies in the waveform model during merger contribute much less to the total SNR.
Note also that, as the hybrids were constructed with the \texttt{TEOBResumS} model as the inspiral
approximant, it is encouraging that we find strong agreement between models with different tidal inspiral approximants.

%%%%%%%%%%%%%%%%%%%%%%%%%%%%%%%%%%%%%%%%%%%%%
\subsection{Importance of NSBH-specific contributions}

The distinguishing difference in the model of an NSBH waveform from a BBH waveform is its behavior close to merger, where strong tidal effects lead to de-phasing of the binary from the standard BBH phase and may lead to disruption of the NS, thereby greatly tapering the amplitude. As the total mass of the NSBH system for this model is expected to be relatively low (not exceeding \(\sim45M_\odot\)), these effects will occur at high frequencies where current ground-based detectors are not highly sensitive. One must then ask how important these effects are to the overall model of the waveform for current and future detectors, and how distinguishable the NSBH-specific effects are from BBH or BNS systems.

To estimate the importance of tidal effects and disruption for the detectability of an \NSBH signal, we compute the SNR at which the NSBH waveform deviates from other waveform approximants covering the parameter space for these merger types; in particular, we compare against both \texttt{PhenomNSBH} with \(\Lambda=0\) to simulate a purely BBH waveform and \texttt{PhenomDNRT}, which contains the same phase model as \texttt{PhenomNSBH} but has a taper applied to the high-frequency merger content of the waveform.

Given an NSBH signal with four internal degrees-of-freedom (\(M,q,\chi,\Lambda\)), the SNR \(\rho\) associated with a 90\% confidence region in parameter space for detection is related to the faithfulness \(\mathcal{F}\) between the NSBH signal (here produced by \texttt{PhenomNSBH}) and another waveform approximant via~\cite{Baird:2012cu}
\beq
\mathcal{F}=1-\frac{3.89}{\rho^2}\,.
\eeq
We initially compute a series of NSBH waveforms using fixed intrinsic parameters \((M_\text{NS},\chi,\Lambda)=(1.35 M_\odot,0,400)\) and allow the mass ratio to vary between 1 and 8. This ensures that we evaluate all merger types captured by the amplitude model in the comparison.

\begin{figure}[ht!]
\includegraphics[width=\columnwidth]{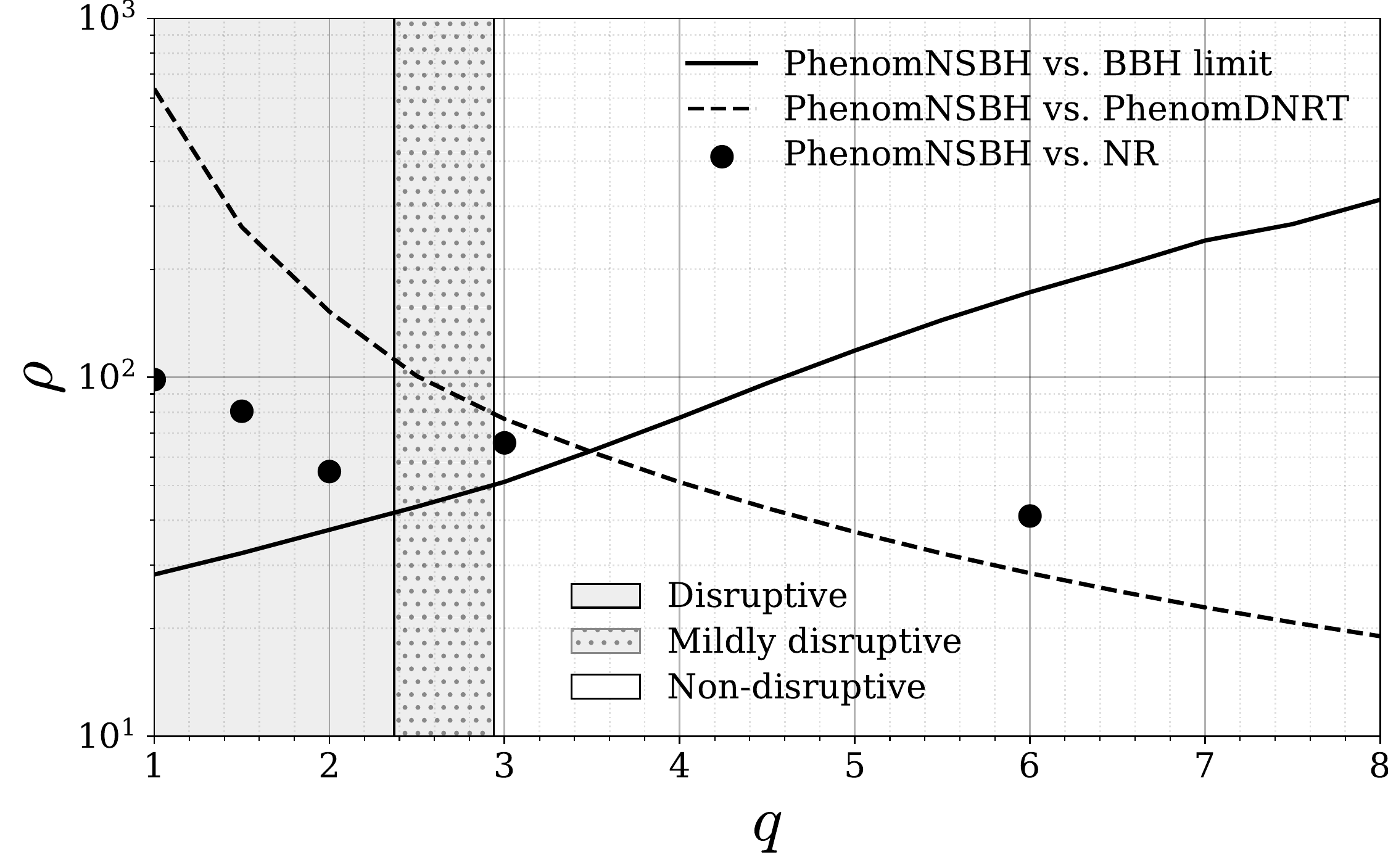}
\caption{The approximate SNR at which the waveforms \texttt{PhenomDNRT} and the BBH-limit of \texttt{PhenomNSBH} become distinguishable from \texttt{PhenomNSBH} is plotted as a function of mass-ratio for a nonspinning NSBH system with tidal deformability \(\Lambda=400\) and NS mass \(1.35 M_\odot\). The shaded regions of the plot indicate different merger types calculated from \texttt{PhenomNSBH}. The solid dots show the SNR computed from mismatches between \texttt{PhenomNSBH} and the NR-hybrid data listed in Table~\ref{T:HybridMatches}. The trends continue to higher mass ratios, where an NSBH signal becomes
effectively indistinguishable from a BBH signal in any realistic detection. The matches between models are computed over the range \([f_1,f_2]=[25,8192]\text{Hz}\) assuming a AZDHP noise curve.
\label{fig:snr-plot}
}
\end{figure}

The SNR resulting from these comparisons is plotted in Fig.~\ref{fig:snr-plot}. Focusing first on the distinguishability SNR between \texttt{PhenomNSBH} and \texttt{PhenomDNRT}, we see that the two models will be easier to distinguish with a modestly loud signal in an Advanced LIGO-type detector as the mass ratio of the system increases.
In the NSBH system, the mass scale is fixed by the NS mass and therefore as \(q\) increases, so too does the total mass \(M\). This increase in \(M\) will push the merger regime of the system into a lower (and more sensitive) frequency band in the detector, making the high-frequency taper applied to the \texttt{NRTidal} model more apparent in the faithfulness calculation. At lower \(q\) in the disruptive regime of the NSBH system, the taper applied to the \texttt{NRTidal} model mimics the disruption at high frequency in the NSBH waveform. Furthermore, these differences between the two models occur at such high frequency that the lack of sensitivity in the detector makes them hard to distinguish.

We stress that this comparison extends the use of \texttt{PhenomDNRT} well beyond the valid parameter space for a BNS system. We wish to test the usefulness of using the \texttt{PhenomDNRT} model to describe an NSBH system, as these systems can share similar amplitude morphologies depending on NSBH merger type. The relatively low distinguishable SNR seen as the mass-ratio increases is not only caused by the change in NSBH morphology but also due to extension of \texttt{PhenomDNRT} beyond its reliable calibration region.

When looking at the comparison between \texttt{PhenomNSBH} with and without tidal effects (\textit{i.e.}, comparing against a BBH waveform), we observe the inverse behavior with changing \(q\).
Even though the disruptive mergers of comparable-mass NSBH binaries lie outside the most sensitive frequency ranges of ground-based detectors, the differences in the waveforms due to tidal effects in the inspiral still allow us to distinguish between BBH and NSBH systems above SNR of 28. This observation is consistent with GW170817~\cite{Abbott:2018wiz} that had an SNR of 32.4 and allowed us to bound the mass-weighted tidal deformability $\tilde{\Lambda}$ away from zero.
As the mass ratio increases, tidal effects scale away as \(q^{-4}\) in the phase and the NSBH signal becomes hard to differentiate from a BBH signal in the non-disruptive regime; the only differences between the two models are the properties of the remnant quantities after merger.

\begin{figure}[ht!]
\subfloat{\includegraphics[width=\columnwidth]{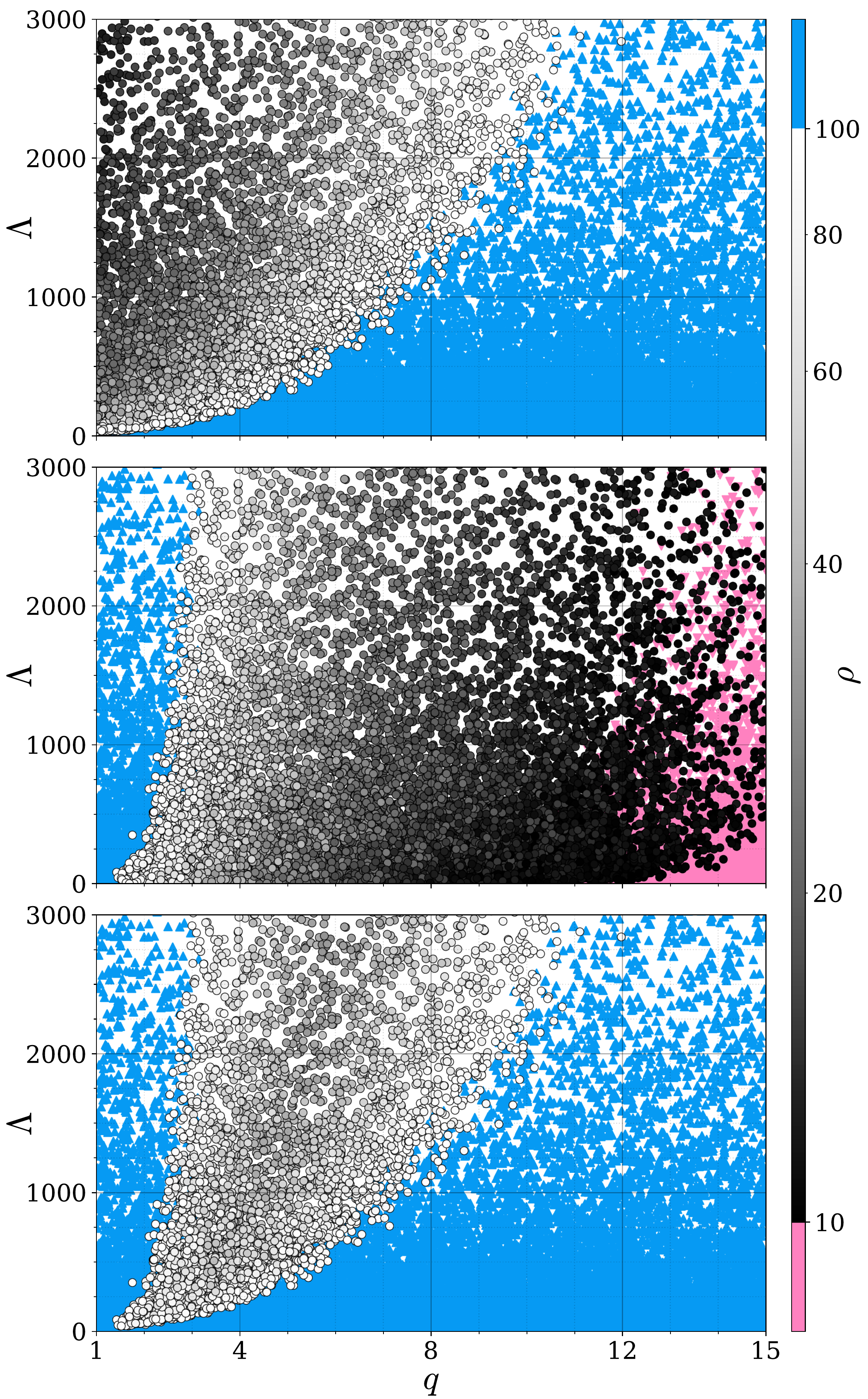}}
\caption{
The approximate SNR at which the waveforms for BNS (\texttt{PhenomDNRT}) and BBH (the BBH-limit of \texttt{PhenomNSBH}) become distinguishable from NSBH (\texttt{PhenomNSBH}), considered over the entire parameter space of \texttt{PhenomNSBH} and projected onto the \(q\)-\(\Lambda\) plane. The top panel displays the distinguishablity of \texttt{PhenomNSBH} from its BBH-limit, the middle panel the distinguishablity of \texttt{PhenomNSBH} from \texttt{PhenomDNRT}, and the bottom panel the maximum distinguishable SNR between \texttt{PhenomNSBH} and the two other models. Distinguishable SNRs below 10 are displayed as pink upside-down triangles and as blue triangles for SNRs above 100.  The AZDHP noise curve is used to compute these results.
\label{fig:SNRdist_scatter}}
\end{figure}

We expand this comparison to include the broader parameter space covered by \texttt{PhenomNSBH}.  Specifically, we assume a AZDHP noise curve and calculate the distinguishability SNR between \texttt{PhenomNSBH} and its BBH limit, and between \texttt{PhenomNSBH} and \texttt{PhenomDNRT} for $\sim 5\times 10^3$ NSBH systems with randomly chosen properties. \(R_{\mathrm{NS}}\) is uniformly sampled between $9\mathrm{km}$ and $16\mathrm{km}$. $M_{\mathrm{NS}}$ is then uniformly sampled over an interval consistent with $\Lambda \in [0, 5000]$ that is dependant on \(R_{\mathrm{NS}}\) and bounded by $1.0\,M_\odot$ and $2.3\,M_\odot$. $\Lambda$ is then calculated from \(R_{\mathrm{NS}}\), $M_{\mathrm{NS}}$ and inverting the universal relation Eq.~(\ref{fig:compactness-fit}). $q$ is uniformly sampled in the interval $[1,15]$ to give $M_{\mathrm{BH}}$, while the BH (aligned) spin is uniformly sampled in the interval $[-0.5, 0.5]$. Our results are collected in Fig.~\ref{fig:SNRdist_scatter}.  The top (middle) panel shows the distinguishability SNR values yielded by \texttt{PhenomNSBH} and its BBH limit (\texttt{PhenomDNRT}), while the bottom panel displays the maximum distinguishable SNR between \texttt{PhenomNSBH} and the two other models.  We see that the general trend described by Fig.~\ref{fig:snr-plot} holds. In particular the characteristic SNR minimum at which the most distinguishable waveform model transitions between \texttt{PhenomDNRT} and the BBH limit persists across parameter space, widening and deepening as tidal deformability increases.

Similar to the model comparisons with \texttt{PhenomNSBH} already presented, the region where \texttt{PhenomNSBH} can be easily distinguished from \texttt{PhenomDNRT} occurs outside of \texttt{PhenomDNRT}'s original parameter bounds and largely results from the extrapolation of the high frequency taper used in the valid region of the BNS model. In contrast \texttt{PhenomNSBH} can be easily distinguished from the BBH limit where both models are valid, and this represents the differences in physical effects modeled by each approximant.

When we consider this transition over the entire parameter space for the model, we find a minimum distinguishable SNR of~27. When constraining \(1.35 M_\odot < M_\text{NS} < 1.4 M_\odot\) we find the minimum distinguishable SNR only increases slightly to~29. Constraining \(\Lambda < 1000\) produces a larger increase in the minimum distinguishable SNR to~35. Applying both cuts in NS mass and \(\Lambda\) increases the minimum distinguishable SNR to~42. These results indicate that the best chance of distinguishing an NSBH signal with current models is from a system with a particularly stiff EOS. It is in this region of relatively low distinguishable SNR that we expect the NSBH model could be most useful. Assuming a single-detector SNR detection threshold of 10, a minimum distinguishable SNR of $\sim 30$ for an optimally-oriented binary system with fixed intrinsic parameters corresponds to a decrease in the distinguishable volume by a factor of $\sim 27$ compared to the detectable volume, and thus roughly one in every 27 NSBH detections of this type could be distinguished from either a BBH or BNS signal.

If a signal were to be detected with $\mathrm{SNR}>60$, comparisons with available NR waveforms suggest that systematic errors in the modeling would enter the waveform and would potentially bias any results inferred from using these models. While we do not anticipate signals with such a high SNR to be seen until third-generation detectors~\cite{Punturo:2010zza,Hild:2010id,Evans:2016mbw} begin operation, should such a signal be detected we will require more accurate NSBH models and potentially more accurate NR simulations of NSBH systems~\cite{Foucart:2018lhe}. However, we have shown that for typical observations we expect either BNS or BBH waveform models to be sufficient.

%%%%%%%%%%%%%%%%%%%%%%%%%%%%%%%%%%%%%%%%%%%%%
\section{Discussion}
\label{sec:discussion}
In this paper we have outlined the construction of \texttt{PhenomNSBH}, an updated waveform model specific to signals from NSBH systems. This model uses an improved amplitude model that identifies distinct merger morphologies and a new tidal phase model, both of which have been calibrated using NR data. The model is valid for systems with mass-ratios ranging from \(q\in[1,15]\) with NS masses between \(M_\text{NS}\in[1,3]M_\odot\), BH spins aligned with the orbital angular momentum ranging between \(\chi\le|0.5|\), and NS tidal deformabilities between \(\Lambda\in[0,5000]\), though we direct the reader toward discussions about untested regions for this model, which can be found at the beginning of Sec.~\ref{sec:analysis}. In addition, the model described here performs well when compared against available NSBH NR waveforms with spinning neutron stars, despite the amplitude model lacking such systems in its calibration.

We have shown in Figs.~\ref{fig:snr-plot} and \ref{fig:SNRdist_scatter} that the NSBH-specific characteristics of \texttt{PhenomNSBH} are distinguishable from other waveform models in different regions of parameter space. As the merger transitions to the non-disruptive regime, the amplitude of the waveform deviates further from a BBH waveform amplitude, which will be distinguishable in ground-based detectors for moderately loud signals. As the merger type becomes less disruptive, the NSBH waveform will easily be distinguishable from a BNS waveform model (\textit{e.g.}, \texttt{PhenomDNRT}) due to the taper at high frequency applied to the latter and lack of ringdown in the signal. The important conclusion to draw from these
results is that for current ground-based detectors, there is only a small region of parameter space where it may be possible to
unambiguously identify an NSBH system given current waveform models. This statement is limited to single observations, and to aligned-spin models that include only
the dominant waveform harmonic.

The waveform model \texttt{PhenomNSBH} described in this paper is an improvement/extension of current NSBH waveform models, but there is certainly room for future advances. While recent cosmological simulations predict that the majority of NSBH systems will have relatively low mass-ratios (\(q\sim[3,5]\))~\cite{10.1093/mnras/sty1613}, even at these low mass-ratios the effects of higher modes~\cite{Varma:2016dnf,Kalaghatgi:2019log} and precession~\cite{Apostolatos:1994mx,Apostolatos:1995pj,Kidder:1993zz} are important to capture the essential physics from the waveform and should be a primary focus of future NSBH waveform modeling efforts. Another avenue for improvement lies in calibrating the phase model against NSBH NR waveforms. These tasks will require a large catalog of new NR simulations at high resolution and spanning a large range of mass-ratios, spins, and tidal deformability.

%%%%%%%%%%%%%%%%%%%%%%%%%%%%%%%%%%%%%%%%%%%%%
\section{Acknowledgments}
The authors would like to express thanks to Frank Ohme, Andrew Matas, and Shrobana Gosh for their work in reviewing the \texttt{LALSuite} implementation of \texttt{PhenomNSBH}, and John Veitch for discussions that initiated this project. The authors would also like to thank the journal referee for insightful  and clarifying comments.
J.T. would like to thank Sarp Akcay for assisting with the production of \texttt{TEOBResumS} waveforms used in the hybrid generation.
T.D. acknowledges support by the European Union’s Horizon 2020 research and innovation program under grant
agreement No 749145, BNSmergers.
J.T. and M.H. were supported by Science and Technology Facilities Council (STFC) grant ST/L000962/1 and thank the Amaldi Research Center for hospitality.
J.T., M.H., S.K., and E.F-J were supported by European Research Council Consolidator Grant 647839.
S.K. acknowledges support by the Max Planck Society’s Independent Research Group Grant.
Analysis and plots in this paper were made using the Python software packages \texttt{LALSuite}~\cite{lalsuite}, \texttt{Matplotlib}~\cite{Hunter:2007}, \texttt{Numpy}~\cite{5725236}, \texttt{PyCBC}~\cite{alex_nitz_2020_3630601}, and \texttt{Scipy}~\cite{4160250}.
The authors are grateful for computational resources provided by the LIGO Laboratory, supported by National Science Foundation Grants PHY-0757058 and PHY-0823459, and by Cardiff University supported by STFC grant ST/I006285/1.

\begin{appendices}

%%%%%%%%%%%%%%%%%%%%%%%%%%%%%%%%%%%%%%%%%%%%%
\section{Amplitude model workflow}
\label{app:amplitude}

For the convenience of the reader, we now outline the construction of the amplitude model in more detail following the flowchart in Sec.~\ref{subsec:amplitude}. To begin, the compactness of the NS is determined from the input tidal deformability, as described in detail in Appendix~\ref{app:EOS}.

We compute the tidal disruption frequency, $f_\text{tide}$, which approximates the frequency at which the external quadrupolar tidal force acting on the NS from the companion BH is comparable in magnitude to the self-gravitating force maintaining the NS. This follows from the initial parameters of the binary according to
\cite{Foucart:2012nc,Shibata:2007zm}
\begin{align}
\label{eq:ftide}
f_{\mathrm{tide}} & = \frac{1}{\pi\left(\chi M_{\mathrm{BH}} + \sqrt{\tilde{r}^3_{\mathrm{tide}}/M_{\mathrm{BH}}}\right)}, \\
\label{eq:rtide}
\tilde{r}_{\mathrm{tide}} & = \xi_{\mathrm{tide}}M_{\mathrm{BH}}\frac{(1-2C)}{\mu},
\end{align}
where $\mu = qC$ and $\xi_{\mathrm{tide}}$ is the largest positive real root of
the following equation,
\begin{multline}
\label{eq:xitide}
0 = \xi_{\mathrm{tide}}^5 - 3\mu\xi_{\mathrm{tide}}^4 + 2\chi\sqrt{\mu^3\xi_{\mathrm{tide}}^{7}} \\
- 3q\xi_{\mathrm{tide}}^2 + 6q\mu\xi_{\mathrm{tide}} - 3q\mu^2\chi^2.
\end{multline}

Next, the ratio of the baryonic mass of the torus remaining after merger to the initial baryonic mass of the NS, $M_{\mathrm{b,torus}}/M_{\mathrm{b,NS}}$, is determined according to fits from \cite{Foucart:2012nc},
\begin{align}
\label{eq:baryonic-mass-ratio}
\frac{M_{\mathrm{b,torus}}}{M_{\mathrm{b,NS}}}
& = 0.296\xi_{\mathrm{tide}}(1-2C) - 0.171qC\bar{r}_{\mathrm{ISCO}},
\end{align}
where $\bar{r}_\mathrm{ISCO}$ is the radius of the innermost stable circular
orbit of a unit-mass BH
\cite{1972ApJ...178..347B},
\begin{align}
\label{eq:risco}
\bar{r}_\mathrm{ISCO} & = \left[
3+Z_2-\mathrm{sign(\chi)}\sqrt{(3-Z_1)(3+Z_1+2Z_2)}\right], \nonumber \\
Z_1 & = 1 + \left(1-\chi^2\right)^{1/3}\left[(1+\chi)^{1/3} + (1-\chi)^{1/3}\right], \nonumber \\
Z_2 & = \sqrt{3\chi^2+Z_1^2}.
\end{align}
The fit for $M_{\mathrm{b,torus}}$ was recently updated in
Ref.~\cite{Foucart:2018rjc}; incorporating it in the amplitude model
would require recalibrating the NSBH amplitude model itself as a whole
and we leave this for future work.

The final mass, $M_f$, and final spin, $\chi_f$, of the remnant BH after merger are
calculated using NSBH-specific fits for the remnant properties parameterized by tidal deformability~\cite{Zappa:2019ntl},
\begin{align}
\label{eq:remnant-model}
F(\eta, \chi, \Lambda) & = F_{\mathrm{BBH}}(\eta, \chi)\frac{
  1 + p_1(\eta, \chi)\Lambda + p_2(\eta, \chi)\Lambda^2
}{\left(1 + [p_3(\eta, \chi)]^{2}\Lambda\right)^2}, \\
\label{eq:remnant-2d-poly}
p_k(\eta, \chi) & = p_{k1}(\chi)\eta + p_{k2}(\chi)\eta^2, \\
\label{eq:remnant-1d-poly}
p_{kj}(\eta, \chi) & = p_{kj0}\chi + p_{kj1}.
\end{align}
The remnant model $F_{\mathrm{BBH}}$ is the model for the final mass and spin
of a BBH coalescence described in \cite{Jimenez-Forteza:2016oae}, and the coefficients $p_{kji}$ for the
final mass $M_f$ and final spin $\chi_f$ can be found in the supplementary material for~\cite{Zappa:2019ntl}.
Once the final mass and spin are determined, we find the ringdown frequency $f_{\mathrm{RD}}$ and quality factor $Q$ via,
\begin{align}
\label{eq:ringdown-frequency}
f_{\mathrm{RD}} & = \frac{\Re{(\tilde{\omega}})}{2 \pi M_f}, \\
\label{eq:quality-factor}
Q & = \frac{\Re{(\tilde{\omega}})}{2\Im{(\tilde{\omega}})},
\end{align}
where \(\tilde{\omega}\) is a fit to the $(l,m,n)=(2,2,0)$ Kerr quasi-normal mode
frequency given in \cite{London:2018nxs},
\begin{align}
\label{eq:omega-tilde}
\tilde{\omega}(\kappa) &= 1.0 + 1.5578e^{2.9031i}\kappa \nonumber\\
& \quad + 1.9510e^{5.9210 i}\kappa^2 + 2.0997e^{2.7606 i}\kappa^3 \nonumber\\
& \quad + 1.4109e^{5.9143i}\kappa^4 + 0.4106e^{2.7952i}\kappa^5 ,\\
\kappa(\chi_f) & = \sqrt{\log_3(2 - \chi_f)}.
\end{align}

The amplitude ansatz in Eq.~\eqref{eq:ampansatz} uses the merger-type-dependent frequencies \(\tilde{f}_0\), \(\tilde{f}_1\), and \(\tilde{f}_2\) to blend the post-Newtonian, pre-merger, and merger-ringdown amplitude contributions together. These frequencies are determined based on the conditions in Table~\ref{tab:merger-dep-quants}. We now list the specific functional form of the various component functions $x_\mathrm{ND}$, $x'_\mathrm{ND}$, $x_\mathrm{D}$ and
$x'_\mathrm{D}$ of the merger-type dependent quantities given in \cite{Pannarale:2015jka}. The non-disruptive fitting functions \(x_\text{ND}\) and \(x'_\text{ND}\) also require the scaled ringdown frequency \(\tilde{f}_{\mathrm{RD}}\) calculated according to
\begin{align}
\label{eq:f_RD_tilde}
\tilde{f}_{\mathrm{RD}} & =
\begin{cases}
0.99 \times 0.98 f_{\mathrm{RD}}, & \Lambda > 1 \\
(1 - 0.02\Lambda + 0.01\Lambda^2) \times 0.98 f_{\mathrm{RD}}, & \Lambda \leq 1,
\end{cases} \\
\label{eq:x_ND}
x_\mathrm{ND} & = \left(\frac{f_{\mathrm{tide}} - \tilde{f}_{\mathrm{RD}}}{\tilde{f}_\mathrm{RD}}\right)^2 - 0.571505C \nonumber \\
& \quad - 0.00508451\chi, \\
\label{eq:x_ND_prime}
x'_\mathrm{ND} & = \left(\frac{f_{\mathrm{tide}} - \tilde{f}_{\mathrm{RD}}}{\tilde{f}_\mathrm{RD}}\right)^2 - 0.657424C \nonumber \\
& \quad - 0.0259977\chi, \\
\label{eq:x_D}
x_\mathrm{D} & = \frac{M_{\mathrm{b,torus}}}{M_{\mathrm{b,NS}}} + 0.424912C \nonumber \\
& \quad + 0.363604\sqrt\eta - 0.060559\chi, \\
\label{eq:x_D_prime}
x'_\mathrm{D} & = \frac{M_{\mathrm{b,torus}}}{M_{\mathrm{b,NS}}} - 0.132754C + 0.576669\sqrt\eta \nonumber \\
& \quad - 0.0603749\chi - 0.0601185\chi^2 \nonumber \\
& \quad - 0.0729134\chi^3.
\end{align}

The amplitude component function for the inspiral, $A_{\mathrm{PN}}$, is
given by the Fourier transform of the time-domain amplitude given in Eq.~(3.14) of \cite{Santamaria:2010yb} using
the stationary phase approximation,
\beq
\label{eq:A_PN}
A_{\mathrm{PN}}(x)  = \sqrt{\frac{2\pi}{3\dot{x}\sqrt x}} 8 \eta x \sqrt{\frac{\pi}{5}} \sum_{k=0}^6\mathcal{A}_k x^{k/2},
\eeq
where \(x=\omega^{2/3}\), \(\omega\) is the orbital angular frequency of the binary, and \(\dot{x}\) is computed using the TaylorT4 expansion~\cite{Buonanno:2002fy}; see \cite{Santamaria:2010yb} for the expansion coefficients $\mathcal{A}_i$.

The phenomenological correction parameter $\gamma'_1$ for the pre-merger region is calculated according to,
\begin{equation}
\label{eq:gamma-1-prime}
\gamma'_1 =
\begin{cases}
1.25, & \Lambda > 1 \\
1 - 0.5\Lambda - 0.25\Lambda^2, & \Lambda \leq 1,
\end{cases}
\end{equation}
where the piecewise definition split at \(\Lambda=1\) is used to smoothly match to the BBH limit where \(\Lambda=0\).

The merger-ringdown component function
$A_{\mathrm{RD}}$ is defined by~\cite{Pannarale:2015jka},
\begin{align}
\label{eq:A_RD}
A_{\mathrm{RD}}(f) &= \epsilon_{\mathrm{tide}} \delta_1 \frac{\sigma^2}{(f-f_{\mathrm{RD}})^2 + \sigma^2/4} f^{-7/6}, \\
\sigma &= \delta'_2 f_{\mathrm{RD}}/Q,
\end{align}
where the phenomenological correction parameter $\delta'_2$ is calculated according to a piecewise definition to smoothly match to the BBH limit as is done for \(\gamma'_1\),
\begin{equation}
\label{eq:delta-2-prime}
\delta'_2= \begin{cases}\displaystyle
\frac{A}{2}\omega^{-}_{x_3,d_3} \left( \frac{f_{\mathrm{tide}} - \tilde{f}_{\mathrm{RD}}}{\tilde{f}_{\mathrm{RD}}} \right), & \Lambda>1 \\
\delta_{2}-2\left(\delta_{2}-b_0\right) \Lambda+\left(\delta_{2}-b_0\right) \Lambda^{2}, & \Lambda \leqslant 1
\end{cases}
\end{equation}
with $A = 1.62496$, $x_3 = 0.0188092$, and $d_3 = 0.338737$, $b_0 = 0.81248$ and $\omega_{f_0,d}^{\pm}(f)$ is a hyperbolic tangent windowing function,
\begin{align}
\label{eq:omega_window}
\omega_{f_0,d}^{\pm}(f) & = \frac{1}{2} \left [ 1 \pm \tanh\left( \frac{4(f-f_0)}{d} \right) \right ].
\end{align}
Note that the
factor of 1/2 multiplying the windowing function $\omega^{-}_{x_3,d_3}$ in Eq.~(\ref{eq:delta-2-prime}) corrects
a typographical error in \cite{Pannarale:2015jka}. The \texttt{PhenomC} phenomenological parameters $\delta_1$, $\delta_2$ and $\gamma_1$ are given as an expansion in symmetric mass-ratio and spins by,
\beq
\label{eq:phenomc-param-model}
\delta_1,\delta_2,\gamma_1 \sim \sum_{i+j\in\{1,2\}} \zeta^{ij}\eta^i\chi^j,
\eeq
with the coefficients \(\zeta^{ij}\) in the $\delta_1$, $\delta_2$, and $\gamma_1$ fit parameters given
in~\cite{Santamaria:2010yb}. We impose the addition constraints that $\delta_1, \gamma_1 \geq 0$ and $\delta_2 \geq 10^{-4}$ to ensure that the amplitude function Eq.~(\ref{eq:ampansatz}) remains positive for all regions of parameter space that \texttt{PhenomNSBH} is expected to be used in. It is necessary to invoke these constraints on these coefficients in the non-spinning limit for $q>25$ and $q>15$ for spinning cases. In this region the model no long remains sensible and comparisons between other BBH waveforms break down. This constraint on the coefficients motivates the suggested upper bound placed on the mass ratio for the parameter space of the model.

%%%%%%%%%%%%%%%%%%%%%%%%%%%%%%%%%%%%%%%%%%%%%
\section{Replacing Equation of State}
\label{app:EOS}

\begin{figure}[ht!]
\includegraphics[width=\columnwidth]{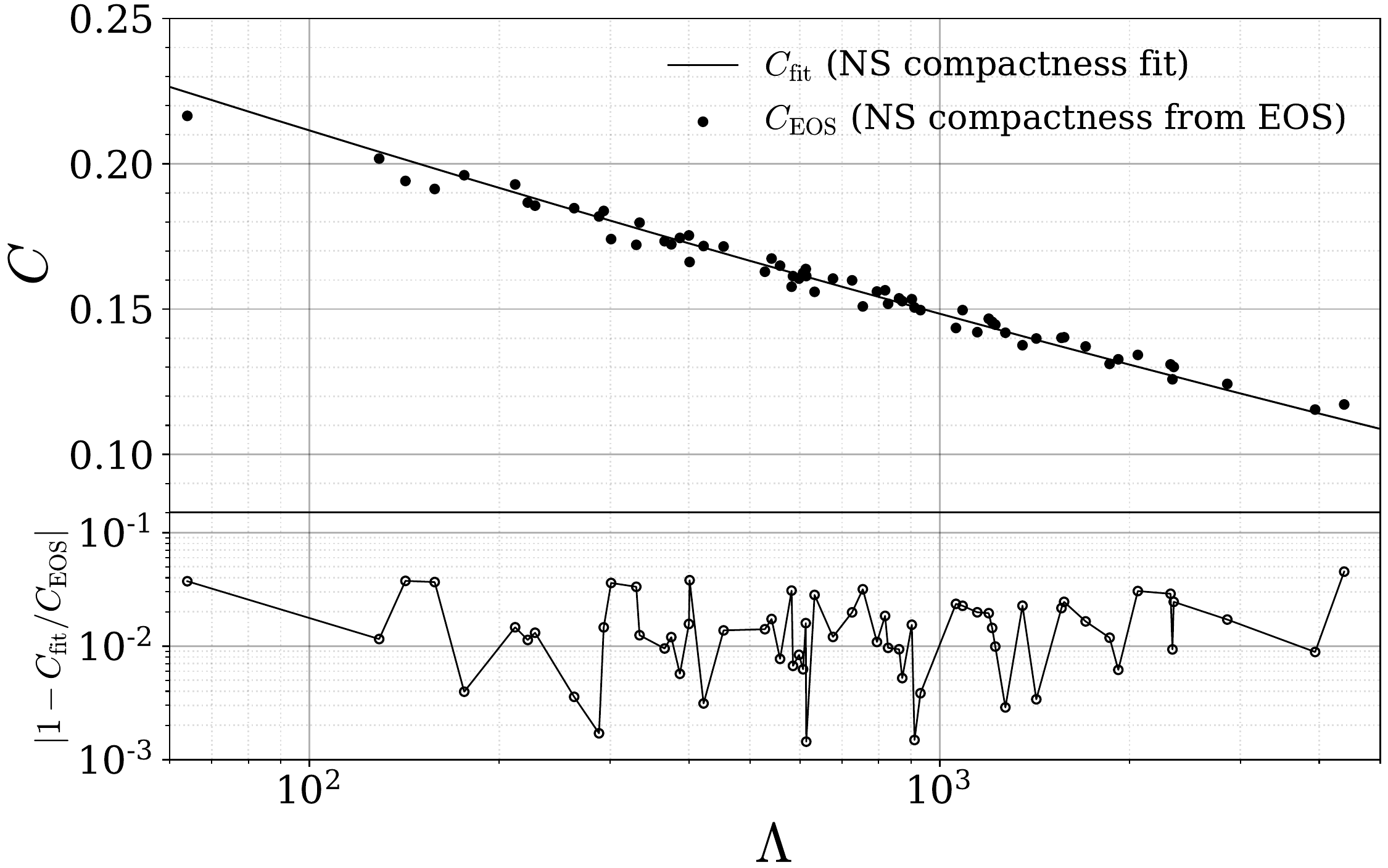}
\caption{Comparison between the NS compactness as calculated from the EOS information presented in \cite{Lackey:2013axa} and the NS compactness fit from \cite{Yagi:2016bkt} which has a root-mean-squared relative percentage error of 1.95\% and maximum relative percentage error of 4.52\%.\label{fig:compactness-fit}}
\end{figure}

Removing explicit EOS-dependence from the \NSBH amplitude model is achieved by finding the compactness \(C\) of the NS from its tidal deformability parameter \(\Lambda\) using the fit determined
in Ref.~\cite{Yagi:2016bkt} with an additional piecewise component for $\Lambda\leq1$ from \cite{SEOBNRNSBH} to smoothly match to the BBH limit,
\beq
\label{eq:compactness}
C(\Lambda) =
\begin{cases}
a_0 + a_1 \log{\Lambda} + a_2 (\log{\Lambda})^2, & \Lambda > 1 \\[8pt]
\begin{aligned}
0.5 + (3a_0&-a_1-1.5)\Lambda^2 \\
&+ (a_1-2a_0+1)\Lambda^3,
\end{aligned} & \Lambda \leq 1,
\end{cases}
\eeq
where $a_0=0.360$, $a_1=-0.0355$, and $a_2=0.000705$.  In Fig.~\ref{fig:compactness-fit} we show how the compactness values yielded by this fit compare to those directly obtained from the EOS information presented in \cite{Lackey:2013axa} by integrating the Tolman-Oppenheimer-Volkoff equations~\cite{Tolman169,PhysRev.55.364,PhysRev.55.374}.

As the original model was calibrated only to a specific set of EOSs, replacing EOS-dependence with the fit in Eq.~\eqref{eq:compactness} will invariable introduce some error to the amplitude model. We conservatively estimate the effects of this error on the model in the following way.

The error in the fit model is given pessimistically as a 6\% error in the computed value of \(C\) across realistic NS EOSs~\cite{Yagi:2016bkt}; for the EOSs used in the calibration of the amplitude model, the error in the fit is bounded by 5\%. We invert the mapping in Eq.~\eqref{eq:compactness} and compute the spread in \(\Lambda\) produced around a given \(\Lambda_0\) by varying the compactness within the 6\% error bounds. We then compute matches across the parameter space of \texttt{PhenomNSBH} between two waveforms with all parameters equal except the tidal deformability, which is fixed at \(\Lambda_0\) for one waveform and allowed to vary between the bounds determined from the compactness error for the other. After sampling waveforms across the model's parameter space, we find a maximum mismatch given by \(\sim10^{-3}\) for the pessimistic 6\% error estimate in the fit.

\end{appendices}

\newpage
\bibliographystyle{apsrev}
\bibliography{paper.bib}

\begin{thebibliography}{81}
\expandafter\ifx\csname natexlab\endcsname\relax\def\natexlab#1{#1}\fi
\expandafter\ifx\csname bibnamefont\endcsname\relax
  \def\bibnamefont#1{#1}\fi
\expandafter\ifx\csname bibfnamefont\endcsname\relax
  \def\bibfnamefont#1{#1}\fi
\expandafter\ifx\csname citenamefont\endcsname\relax
  \def\citenamefont#1{#1}\fi
\expandafter\ifx\csname url\endcsname\relax
  \def\url#1{\texttt{#1}}\fi
\expandafter\ifx\csname urlprefix\endcsname\relax\def\urlprefix{URL }\fi
\providecommand{\bibinfo}[2]{#2}
\providecommand{\eprint}[2][]{\url{#2}}

\bibitem[{\citenamefont{Aasi et~al.}(2015)}]{TheLIGOScientific:2014jea}
\bibinfo{author}{\bibfnamefont{J.}~\bibnamefont{Aasi}} \bibnamefont{et~al.}
  (\bibinfo{collaboration}{LIGO Scientific}), \bibinfo{journal}{Class. Quant.
  Grav.} \textbf{\bibinfo{volume}{32}}, \bibinfo{pages}{074001}
  (\bibinfo{year}{2015}), \eprint{1411.4547}.

\bibitem[{\citenamefont{Acernese et~al.}(2015)}]{TheVirgo:2014hva}
\bibinfo{author}{\bibfnamefont{F.}~\bibnamefont{Acernese}} \bibnamefont{et~al.}
  (\bibinfo{collaboration}{VIRGO}), \bibinfo{journal}{Class. Quant. Grav.}
  \textbf{\bibinfo{volume}{32}}, \bibinfo{pages}{024001}
  (\bibinfo{year}{2015}), \eprint{1408.3978}.

\bibitem[{\citenamefont{Vallisneri et~al.}(2015)\citenamefont{Vallisneri,
  Kanner, Williams, Weinstein, and Stephens}}]{Vallisneri:2014vxa}
\bibinfo{author}{\bibfnamefont{M.}~\bibnamefont{Vallisneri}},
  \bibinfo{author}{\bibfnamefont{J.}~\bibnamefont{Kanner}},
  \bibinfo{author}{\bibfnamefont{R.}~\bibnamefont{Williams}},
  \bibinfo{author}{\bibfnamefont{A.}~\bibnamefont{Weinstein}},
  \bibnamefont{and} \bibinfo{author}{\bibfnamefont{B.}~\bibnamefont{Stephens}},
  \bibinfo{journal}{J. Phys. Conf. Ser.} \textbf{\bibinfo{volume}{610}},
  \bibinfo{pages}{012021} (\bibinfo{year}{2015}), \eprint{1410.4839}.

\bibitem[{\citenamefont{Collaboration et~al.}(2019)\citenamefont{Collaboration,
  the Virgo~Collaboration, Abbott et~al.}}]{collaboration2019open}
\bibinfo{author}{\bibfnamefont{T.~L.~S.} \bibnamefont{Collaboration}},
  \bibinfo{author}{\bibnamefont{the Virgo~Collaboration}},
  \bibinfo{author}{\bibfnamefont{R.}~\bibnamefont{Abbott}},
  \bibnamefont{et~al.}, \emph{\bibinfo{title}{Open data from the first and
  second observing runs of advanced ligo and advanced virgo}}
  (\bibinfo{year}{2019}), \eprint{1912.11716}.

\bibitem[{\citenamefont{Abbott
  et~al.}(2019{\natexlab{a}})}]{LIGOScientific:2018mvr}
\bibinfo{author}{\bibfnamefont{B.~P.} \bibnamefont{Abbott}}
  \bibnamefont{et~al.} (\bibinfo{collaboration}{LIGO Scientific, Virgo}),
  \bibinfo{journal}{Phys. Rev.} \textbf{\bibinfo{volume}{X9}},
  \bibinfo{pages}{031040} (\bibinfo{year}{2019}{\natexlab{a}}),
  \eprint{1811.12907}.

\bibitem[{\citenamefont{Venumadhav et~al.}(2019)\citenamefont{Venumadhav,
  Zackay, Roulet, Dai, and Zaldarriaga}}]{Venumadhav:2019lyq}
\bibinfo{author}{\bibfnamefont{T.}~\bibnamefont{Venumadhav}},
  \bibinfo{author}{\bibfnamefont{B.}~\bibnamefont{Zackay}},
  \bibinfo{author}{\bibfnamefont{J.}~\bibnamefont{Roulet}},
  \bibinfo{author}{\bibfnamefont{L.}~\bibnamefont{Dai}}, \bibnamefont{and}
  \bibinfo{author}{\bibfnamefont{M.}~\bibnamefont{Zaldarriaga}}
  (\bibinfo{year}{2019}), \eprint{1904.07214}.

\bibitem[{\citenamefont{Nitz et~al.}(2019{\natexlab{a}})\citenamefont{Nitz,
  Dent, Davies, Kumar, Capano, Harry, Mozzon, Nuttall, Lundgren, and
  Tápai}}]{Nitz:2019hdf}
\bibinfo{author}{\bibfnamefont{A.~H.} \bibnamefont{Nitz}},
  \bibinfo{author}{\bibfnamefont{T.}~\bibnamefont{Dent}},
  \bibinfo{author}{\bibfnamefont{G.~S.} \bibnamefont{Davies}},
  \bibinfo{author}{\bibfnamefont{S.}~\bibnamefont{Kumar}},
  \bibinfo{author}{\bibfnamefont{C.~D.} \bibnamefont{Capano}},
  \bibinfo{author}{\bibfnamefont{I.}~\bibnamefont{Harry}},
  \bibinfo{author}{\bibfnamefont{S.}~\bibnamefont{Mozzon}},
  \bibinfo{author}{\bibfnamefont{L.}~\bibnamefont{Nuttall}},
  \bibinfo{author}{\bibfnamefont{A.}~\bibnamefont{Lundgren}}, \bibnamefont{and}
  \bibinfo{author}{\bibfnamefont{M.}~\bibnamefont{Tápai}}
  (\bibinfo{year}{2019}{\natexlab{a}}), \eprint{1910.05331}.

\bibitem[{\citenamefont{Nitz et~al.}(2019{\natexlab{b}})\citenamefont{Nitz,
  Nielsen, and Capano}}]{Nitz:2019bxt}
\bibinfo{author}{\bibfnamefont{A.~H.} \bibnamefont{Nitz}},
  \bibinfo{author}{\bibfnamefont{A.~B.} \bibnamefont{Nielsen}},
  \bibnamefont{and} \bibinfo{author}{\bibfnamefont{C.~D.}
  \bibnamefont{Capano}}, \bibinfo{journal}{Astrophys. J.}
  \textbf{\bibinfo{volume}{876}}, \bibinfo{pages}{L4}
  (\bibinfo{year}{2019}{\natexlab{b}}), \bibinfo{note}{[Astrophys. J.
  Lett.876,L4(2019)]}, \eprint{1902.09496}.

\bibitem[{\citenamefont{Abbott et~al.}(2019{\natexlab{b}})}]{Abbott:2018wiz}
\bibinfo{author}{\bibfnamefont{B.~P.} \bibnamefont{Abbott}}
  \bibnamefont{et~al.} (\bibinfo{collaboration}{LIGO Scientific, Virgo}),
  \bibinfo{journal}{Phys. Rev.} \textbf{\bibinfo{volume}{X9}},
  \bibinfo{pages}{011001} (\bibinfo{year}{2019}{\natexlab{b}}),
  \eprint{1805.11579}.

\bibitem[{\citenamefont{Abbott et~al.}(2020)}]{Abbott:2020uma}
\bibinfo{author}{\bibfnamefont{B.~P.} \bibnamefont{Abbott}}
  \bibnamefont{et~al.} (\bibinfo{collaboration}{LIGO Scientific, Virgo})
  (\bibinfo{year}{2020}), \eprint{2001.01761}.

\bibitem[{\citenamefont{Coughlin and Dietrich}(2019)}]{Coughlin:2019kqf}
\bibinfo{author}{\bibfnamefont{M.~W.} \bibnamefont{Coughlin}} \bibnamefont{and}
  \bibinfo{author}{\bibfnamefont{T.}~\bibnamefont{Dietrich}},
  \bibinfo{journal}{Phys. Rev.} \textbf{\bibinfo{volume}{D100}},
  \bibinfo{pages}{043011} (\bibinfo{year}{2019}), \eprint{1901.06052}.

\bibitem[{\citenamefont{Kyutoku et~al.}(2020)\citenamefont{Kyutoku,
  Fujibayashi, Hayashi, Kawaguchi, Kiuchi, Shibata, and
  Tanaka}}]{Kyutoku:2020xka}
\bibinfo{author}{\bibfnamefont{K.}~\bibnamefont{Kyutoku}},
  \bibinfo{author}{\bibfnamefont{S.}~\bibnamefont{Fujibayashi}},
  \bibinfo{author}{\bibfnamefont{K.}~\bibnamefont{Hayashi}},
  \bibinfo{author}{\bibfnamefont{K.}~\bibnamefont{Kawaguchi}},
  \bibinfo{author}{\bibfnamefont{K.}~\bibnamefont{Kiuchi}},
  \bibinfo{author}{\bibfnamefont{M.}~\bibnamefont{Shibata}}, \bibnamefont{and}
  \bibinfo{author}{\bibfnamefont{M.}~\bibnamefont{Tanaka}},
  \bibinfo{journal}{Astrophys. J.} \textbf{\bibinfo{volume}{890}},
  \bibinfo{pages}{L4} (\bibinfo{year}{2020}), \eprint{2001.04474}.

\bibitem[{\citenamefont{Cotesta et~al.}(2018)\citenamefont{Cotesta, Buonanno,
  Bohé, Taracchini, Hinder, and Ossokine}}]{Cotesta:2018fcv}
\bibinfo{author}{\bibfnamefont{R.}~\bibnamefont{Cotesta}},
  \bibinfo{author}{\bibfnamefont{A.}~\bibnamefont{Buonanno}},
  \bibinfo{author}{\bibfnamefont{A.}~\bibnamefont{Bohé}},
  \bibinfo{author}{\bibfnamefont{A.}~\bibnamefont{Taracchini}},
  \bibinfo{author}{\bibfnamefont{I.}~\bibnamefont{Hinder}}, \bibnamefont{and}
  \bibinfo{author}{\bibfnamefont{S.}~\bibnamefont{Ossokine}},
  \bibinfo{journal}{Phys. Rev.} \textbf{\bibinfo{volume}{D98}},
  \bibinfo{pages}{084028} (\bibinfo{year}{2018}), \eprint{1803.10701}.

\bibitem[{\citenamefont{Khan et~al.}(2019)\citenamefont{Khan, Chatziioannou,
  Hannam, and Ohme}}]{PhysRevD.100.024059}
\bibinfo{author}{\bibfnamefont{S.}~\bibnamefont{Khan}},
  \bibinfo{author}{\bibfnamefont{K.}~\bibnamefont{Chatziioannou}},
  \bibinfo{author}{\bibfnamefont{M.}~\bibnamefont{Hannam}}, \bibnamefont{and}
  \bibinfo{author}{\bibfnamefont{F.}~\bibnamefont{Ohme}},
  \bibinfo{journal}{Phys. Rev. D} \textbf{\bibinfo{volume}{100}},
  \bibinfo{pages}{024059} (\bibinfo{year}{2019}),
  \urlprefix\url{https://link.aps.org/doi/10.1103/PhysRevD.100.024059}.

\bibitem[{\citenamefont{Khan et~al.}(2020)\citenamefont{Khan, Ohme,
  Chatziioannou, and Hannam}}]{Khan:2019kot}
\bibinfo{author}{\bibfnamefont{S.}~\bibnamefont{Khan}},
  \bibinfo{author}{\bibfnamefont{F.}~\bibnamefont{Ohme}},
  \bibinfo{author}{\bibfnamefont{K.}~\bibnamefont{Chatziioannou}},
  \bibnamefont{and} \bibinfo{author}{\bibfnamefont{M.}~\bibnamefont{Hannam}},
  \bibinfo{journal}{Phys. Rev.} \textbf{\bibinfo{volume}{D101}},
  \bibinfo{pages}{024056} (\bibinfo{year}{2020}), \eprint{1911.06050}.

\bibitem[{\citenamefont{Varma et~al.}(2019{\natexlab{a}})\citenamefont{Varma,
  Field, Scheel, Blackman, Gerosa, Stein, Kidder, and
  Pfeiffer}}]{Varma:2019csw}
\bibinfo{author}{\bibfnamefont{V.}~\bibnamefont{Varma}},
  \bibinfo{author}{\bibfnamefont{S.~E.} \bibnamefont{Field}},
  \bibinfo{author}{\bibfnamefont{M.~A.} \bibnamefont{Scheel}},
  \bibinfo{author}{\bibfnamefont{J.}~\bibnamefont{Blackman}},
  \bibinfo{author}{\bibfnamefont{D.}~\bibnamefont{Gerosa}},
  \bibinfo{author}{\bibfnamefont{L.~C.} \bibnamefont{Stein}},
  \bibinfo{author}{\bibfnamefont{L.~E.} \bibnamefont{Kidder}},
  \bibnamefont{and} \bibinfo{author}{\bibfnamefont{H.~P.}
  \bibnamefont{Pfeiffer}}, \bibinfo{journal}{Phys. Rev. Research.}
  \textbf{\bibinfo{volume}{1}}, \bibinfo{pages}{033015}
  (\bibinfo{year}{2019}{\natexlab{a}}), \eprint{1905.09300}.

\bibitem[{\citenamefont{Varma et~al.}(2019{\natexlab{b}})\citenamefont{Varma,
  Field, Scheel, Blackman, Kidder, and Pfeiffer}}]{PhysRevD.99.064045}
\bibinfo{author}{\bibfnamefont{V.}~\bibnamefont{Varma}},
  \bibinfo{author}{\bibfnamefont{S.~E.} \bibnamefont{Field}},
  \bibinfo{author}{\bibfnamefont{M.~A.} \bibnamefont{Scheel}},
  \bibinfo{author}{\bibfnamefont{J.}~\bibnamefont{Blackman}},
  \bibinfo{author}{\bibfnamefont{L.~E.} \bibnamefont{Kidder}},
  \bibnamefont{and} \bibinfo{author}{\bibfnamefont{H.~P.}
  \bibnamefont{Pfeiffer}}, \bibinfo{journal}{Phys. Rev. D}
  \textbf{\bibinfo{volume}{99}}, \bibinfo{pages}{064045}
  (\bibinfo{year}{2019}{\natexlab{b}}),
  \urlprefix\url{https://link.aps.org/doi/10.1103/PhysRevD.99.064045}.

\bibitem[{\citenamefont{Flanagan and Hinderer}(2008)}]{Flanagan:2007ix}
\bibinfo{author}{\bibfnamefont{{\'E}.~{\'E}.} \bibnamefont{Flanagan}}
  \bibnamefont{and} \bibinfo{author}{\bibfnamefont{T.}~\bibnamefont{Hinderer}},
  \bibinfo{journal}{Phys. Rev. D} \textbf{\bibinfo{volume}{77}},
  \bibinfo{pages}{021502(R)} (\bibinfo{year}{2008}), \eprint{0709.1915}.

\bibitem[{\citenamefont{Nagar et~al.}(2018)}]{Nagar:2018zoe}
\bibinfo{author}{\bibfnamefont{A.}~\bibnamefont{Nagar}} \bibnamefont{et~al.},
  \bibinfo{journal}{Phys. Rev.} \textbf{\bibinfo{volume}{D98}},
  \bibinfo{pages}{104052} (\bibinfo{year}{2018}), \eprint{1806.01772}.

\bibitem[{\citenamefont{Hinderer et~al.}(2016)\citenamefont{Hinderer,
  Taracchini, Foucart, Buonanno, Steinhoff, Duez, Kidder, Pfeiffer, Scheel,
  Szilagyi et~al.}}]{PhysRevLett.116.181101}
\bibinfo{author}{\bibfnamefont{T.}~\bibnamefont{Hinderer}},
  \bibinfo{author}{\bibfnamefont{A.}~\bibnamefont{Taracchini}},
  \bibinfo{author}{\bibfnamefont{F.}~\bibnamefont{Foucart}},
  \bibinfo{author}{\bibfnamefont{A.}~\bibnamefont{Buonanno}},
  \bibinfo{author}{\bibfnamefont{J.}~\bibnamefont{Steinhoff}},
  \bibinfo{author}{\bibfnamefont{M.}~\bibnamefont{Duez}},
  \bibinfo{author}{\bibfnamefont{L.~E.} \bibnamefont{Kidder}},
  \bibinfo{author}{\bibfnamefont{H.~P.} \bibnamefont{Pfeiffer}},
  \bibinfo{author}{\bibfnamefont{M.~A.} \bibnamefont{Scheel}},
  \bibinfo{author}{\bibfnamefont{B.}~\bibnamefont{Szilagyi}},
  \bibnamefont{et~al.}, \bibinfo{journal}{Phys. Rev. Lett.}
  \textbf{\bibinfo{volume}{116}}, \bibinfo{pages}{181101}
  (\bibinfo{year}{2016}),
  \urlprefix\url{https://link.aps.org/doi/10.1103/PhysRevLett.116.181101}.

\bibitem[{\citenamefont{Steinhoff et~al.}(2016)\citenamefont{Steinhoff,
  Hinderer, Buonanno, and Taracchini}}]{PhysRevD.94.104028}
\bibinfo{author}{\bibfnamefont{J.}~\bibnamefont{Steinhoff}},
  \bibinfo{author}{\bibfnamefont{T.}~\bibnamefont{Hinderer}},
  \bibinfo{author}{\bibfnamefont{A.}~\bibnamefont{Buonanno}}, \bibnamefont{and}
  \bibinfo{author}{\bibfnamefont{A.}~\bibnamefont{Taracchini}},
  \bibinfo{journal}{Phys. Rev. D} \textbf{\bibinfo{volume}{94}},
  \bibinfo{pages}{104028} (\bibinfo{year}{2016}),
  \urlprefix\url{https://link.aps.org/doi/10.1103/PhysRevD.94.104028}.

\bibitem[{\citenamefont{Lackey et~al.}(2019)\citenamefont{Lackey, P\"urrer,
  Taracchini, and Marsat}}]{PhysRevD.100.024002}
\bibinfo{author}{\bibfnamefont{B.~D.} \bibnamefont{Lackey}},
  \bibinfo{author}{\bibfnamefont{M.}~\bibnamefont{P\"urrer}},
  \bibinfo{author}{\bibfnamefont{A.}~\bibnamefont{Taracchini}},
  \bibnamefont{and} \bibinfo{author}{\bibfnamefont{S.}~\bibnamefont{Marsat}},
  \bibinfo{journal}{Phys. Rev. D} \textbf{\bibinfo{volume}{100}},
  \bibinfo{pages}{024002} (\bibinfo{year}{2019}),
  \urlprefix\url{https://link.aps.org/doi/10.1103/PhysRevD.100.024002}.

\bibitem[{\citenamefont{Dietrich et~al.}(2017)\citenamefont{Dietrich, Bernuzzi,
  and Tichy}}]{Dietrich:2017aum}
\bibinfo{author}{\bibfnamefont{T.}~\bibnamefont{Dietrich}},
  \bibinfo{author}{\bibfnamefont{S.}~\bibnamefont{Bernuzzi}}, \bibnamefont{and}
  \bibinfo{author}{\bibfnamefont{W.}~\bibnamefont{Tichy}},
  \bibinfo{journal}{Phys. Rev. D} \textbf{\bibinfo{volume}{96}},
  \bibinfo{pages}{121501(R)} (\bibinfo{year}{2017}), \eprint{1706.02969}.

\bibitem[{\citenamefont{Dietrich
  et~al.}(2019{\natexlab{a}})}]{Dietrich:2018uni}
\bibinfo{author}{\bibfnamefont{T.}~\bibnamefont{Dietrich}}
  \bibnamefont{et~al.}, \bibinfo{journal}{Phys. Rev.}
  \textbf{\bibinfo{volume}{D99}}, \bibinfo{pages}{024029}
  (\bibinfo{year}{2019}{\natexlab{a}}), \eprint{1804.02235}.

\bibitem[{\citenamefont{Dietrich
  et~al.}(2019{\natexlab{b}})\citenamefont{Dietrich, Samajdar, Khan,
  Johnson-McDaniel, Dudi, and Tichy}}]{Dietrich:2019kaq}
\bibinfo{author}{\bibfnamefont{T.}~\bibnamefont{Dietrich}},
  \bibinfo{author}{\bibfnamefont{A.}~\bibnamefont{Samajdar}},
  \bibinfo{author}{\bibfnamefont{S.}~\bibnamefont{Khan}},
  \bibinfo{author}{\bibfnamefont{N.~K.} \bibnamefont{Johnson-McDaniel}},
  \bibinfo{author}{\bibfnamefont{R.}~\bibnamefont{Dudi}}, \bibnamefont{and}
  \bibinfo{author}{\bibfnamefont{W.}~\bibnamefont{Tichy}}
  (\bibinfo{year}{2019}{\natexlab{b}}), \eprint{1905.06011}.

\bibitem[{\citenamefont{Clark et~al.}(2016)\citenamefont{Clark, Bauswein,
  Stergioulas, and Shoemaker}}]{Clark:2015zxa}
\bibinfo{author}{\bibfnamefont{J.~A.} \bibnamefont{Clark}},
  \bibinfo{author}{\bibfnamefont{A.}~\bibnamefont{Bauswein}},
  \bibinfo{author}{\bibfnamefont{N.}~\bibnamefont{Stergioulas}},
  \bibnamefont{and}
  \bibinfo{author}{\bibfnamefont{D.}~\bibnamefont{Shoemaker}},
  \bibinfo{journal}{Class. Quant. Grav.} \textbf{\bibinfo{volume}{33}},
  \bibinfo{pages}{085003} (\bibinfo{year}{2016}), \eprint{1509.08522}.

\bibitem[{\citenamefont{Tsang et~al.}(2019)\citenamefont{Tsang, Dietrich, and
  Van Den~Broeck}}]{PhysRevD.100.044047}
\bibinfo{author}{\bibfnamefont{K.~W.} \bibnamefont{Tsang}},
  \bibinfo{author}{\bibfnamefont{T.}~\bibnamefont{Dietrich}}, \bibnamefont{and}
  \bibinfo{author}{\bibfnamefont{C.}~\bibnamefont{Van Den~Broeck}},
  \bibinfo{journal}{Phys. Rev. D} \textbf{\bibinfo{volume}{100}},
  \bibinfo{pages}{044047} (\bibinfo{year}{2019}),
  \urlprefix\url{https://link.aps.org/doi/10.1103/PhysRevD.100.044047}.

\bibitem[{\citenamefont{Breschi et~al.}(2019)\citenamefont{Breschi, Bernuzzi,
  Zappa, Agathos, Perego, Radice, and Nagar}}]{PhysRevD.100.104029}
\bibinfo{author}{\bibfnamefont{M.}~\bibnamefont{Breschi}},
  \bibinfo{author}{\bibfnamefont{S.}~\bibnamefont{Bernuzzi}},
  \bibinfo{author}{\bibfnamefont{F.}~\bibnamefont{Zappa}},
  \bibinfo{author}{\bibfnamefont{M.}~\bibnamefont{Agathos}},
  \bibinfo{author}{\bibfnamefont{A.}~\bibnamefont{Perego}},
  \bibinfo{author}{\bibfnamefont{D.}~\bibnamefont{Radice}}, \bibnamefont{and}
  \bibinfo{author}{\bibfnamefont{A.}~\bibnamefont{Nagar}},
  \bibinfo{journal}{Phys. Rev. D} \textbf{\bibinfo{volume}{100}},
  \bibinfo{pages}{104029} (\bibinfo{year}{2019}),
  \urlprefix\url{https://link.aps.org/doi/10.1103/PhysRevD.100.104029}.

\bibitem[{\citenamefont{Foucart et~al.}(2019)}]{Foucart:2018lhe}
\bibinfo{author}{\bibfnamefont{F.}~\bibnamefont{Foucart}} \bibnamefont{et~al.},
  \bibinfo{journal}{Phys. Rev.} \textbf{\bibinfo{volume}{D99}},
  \bibinfo{pages}{044008} (\bibinfo{year}{2019}), \eprint{1812.06988}.

\bibitem[{\citenamefont{Dudi et~al.}(2018)\citenamefont{Dudi, Pannarale,
  Dietrich, Hannam, Bernuzzi, Ohme, and Brügmann}}]{Dudi:2018jzn}
\bibinfo{author}{\bibfnamefont{R.}~\bibnamefont{Dudi}},
  \bibinfo{author}{\bibfnamefont{F.}~\bibnamefont{Pannarale}},
  \bibinfo{author}{\bibfnamefont{T.}~\bibnamefont{Dietrich}},
  \bibinfo{author}{\bibfnamefont{M.}~\bibnamefont{Hannam}},
  \bibinfo{author}{\bibfnamefont{S.}~\bibnamefont{Bernuzzi}},
  \bibinfo{author}{\bibfnamefont{F.}~\bibnamefont{Ohme}}, \bibnamefont{and}
  \bibinfo{author}{\bibfnamefont{B.}~\bibnamefont{Brügmann}},
  \bibinfo{journal}{Phys. Rev.} \textbf{\bibinfo{volume}{D98}},
  \bibinfo{pages}{084061} (\bibinfo{year}{2018}), \eprint{1808.09749}.

\bibitem[{\citenamefont{Pannarale et~al.}(2011)\citenamefont{Pannarale,
  Rezzolla, Ohme, and Read}}]{Pannarale:2011pk}
\bibinfo{author}{\bibfnamefont{F.}~\bibnamefont{Pannarale}},
  \bibinfo{author}{\bibfnamefont{L.}~\bibnamefont{Rezzolla}},
  \bibinfo{author}{\bibfnamefont{F.}~\bibnamefont{Ohme}}, \bibnamefont{and}
  \bibinfo{author}{\bibfnamefont{J.~S.} \bibnamefont{Read}},
  \bibinfo{journal}{Phys. Rev.} \textbf{\bibinfo{volume}{D84}},
  \bibinfo{pages}{104017} (\bibinfo{year}{2011}), \eprint{1103.3526}.

\bibitem[{\citenamefont{Yamamoto et~al.}(2008)\citenamefont{Yamamoto, Shibata,
  and Taniguchi}}]{PhysRevD.78.064054}
\bibinfo{author}{\bibfnamefont{T.}~\bibnamefont{Yamamoto}},
  \bibinfo{author}{\bibfnamefont{M.}~\bibnamefont{Shibata}}, \bibnamefont{and}
  \bibinfo{author}{\bibfnamefont{K.}~\bibnamefont{Taniguchi}},
  \bibinfo{journal}{Phys. Rev. D} \textbf{\bibinfo{volume}{78}},
  \bibinfo{pages}{064054} (\bibinfo{year}{2008}),
  \urlprefix\url{https://link.aps.org/doi/10.1103/PhysRevD.78.064054}.

\bibitem[{\citenamefont{Foucart et~al.}(2013)\citenamefont{Foucart, Buchman,
  Duez, Grudich, Kidder, MacDonald, Mroue, Pfeiffer, Scheel, and
  Szilagyi}}]{Foucart:2013psa}
\bibinfo{author}{\bibfnamefont{F.}~\bibnamefont{Foucart}},
  \bibinfo{author}{\bibfnamefont{L.}~\bibnamefont{Buchman}},
  \bibinfo{author}{\bibfnamefont{M.~D.} \bibnamefont{Duez}},
  \bibinfo{author}{\bibfnamefont{M.}~\bibnamefont{Grudich}},
  \bibinfo{author}{\bibfnamefont{L.~E.} \bibnamefont{Kidder}},
  \bibinfo{author}{\bibfnamefont{I.}~\bibnamefont{MacDonald}},
  \bibinfo{author}{\bibfnamefont{A.}~\bibnamefont{Mroue}},
  \bibinfo{author}{\bibfnamefont{H.~P.} \bibnamefont{Pfeiffer}},
  \bibinfo{author}{\bibfnamefont{M.~A.} \bibnamefont{Scheel}},
  \bibnamefont{and} \bibinfo{author}{\bibfnamefont{B.}~\bibnamefont{Szilagyi}},
  \bibinfo{journal}{Phys. Rev.} \textbf{\bibinfo{volume}{D88}},
  \bibinfo{pages}{064017} (\bibinfo{year}{2013}), \eprint{1307.7685}.

\bibitem[{\citenamefont{Kyutoku et~al.}(2011)\citenamefont{Kyutoku, Okawa,
  Shibata, and Taniguchi}}]{Kyutoku:2011vz}
\bibinfo{author}{\bibfnamefont{K.}~\bibnamefont{Kyutoku}},
  \bibinfo{author}{\bibfnamefont{H.}~\bibnamefont{Okawa}},
  \bibinfo{author}{\bibfnamefont{M.}~\bibnamefont{Shibata}}, \bibnamefont{and}
  \bibinfo{author}{\bibfnamefont{K.}~\bibnamefont{Taniguchi}},
  \bibinfo{journal}{Phys. Rev.} \textbf{\bibinfo{volume}{D84}},
  \bibinfo{pages}{064018} (\bibinfo{year}{2011}), \eprint{1108.1189}.

\bibitem[{\citenamefont{Kyutoku et~al.}(2010)\citenamefont{Kyutoku, Shibata,
  and Taniguchi}}]{Kyutoku:2010zd}
\bibinfo{author}{\bibfnamefont{K.}~\bibnamefont{Kyutoku}},
  \bibinfo{author}{\bibfnamefont{M.}~\bibnamefont{Shibata}}, \bibnamefont{and}
  \bibinfo{author}{\bibfnamefont{K.}~\bibnamefont{Taniguchi}},
  \bibinfo{journal}{Phys. Rev.} \textbf{\bibinfo{volume}{D82}},
  \bibinfo{pages}{044049} (\bibinfo{year}{2010}), \bibinfo{note}{[Erratum:
  Phys. Rev.D84,049902(2011)]}, \eprint{1008.1460}.

\bibitem[{\citenamefont{Pannarale
  et~al.}(2015{\natexlab{a}})\citenamefont{Pannarale, Berti, Kyutoku, Lackey,
  and Shibata}}]{Pannarale:2015jia}
\bibinfo{author}{\bibfnamefont{F.}~\bibnamefont{Pannarale}},
  \bibinfo{author}{\bibfnamefont{E.}~\bibnamefont{Berti}},
  \bibinfo{author}{\bibfnamefont{K.}~\bibnamefont{Kyutoku}},
  \bibinfo{author}{\bibfnamefont{B.~D.} \bibnamefont{Lackey}},
  \bibnamefont{and} \bibinfo{author}{\bibfnamefont{M.}~\bibnamefont{Shibata}},
  \bibinfo{journal}{Phys. Rev.} \textbf{\bibinfo{volume}{D92}},
  \bibinfo{pages}{081504} (\bibinfo{year}{2015}{\natexlab{a}}),
  \eprint{1509.06209}.

\bibitem[{\citenamefont{Lackey et~al.}(2014)\citenamefont{Lackey, Kyutoku,
  Shibata, Brady, and Friedman}}]{Lackey:2013axa}
\bibinfo{author}{\bibfnamefont{B.~D.} \bibnamefont{Lackey}},
  \bibinfo{author}{\bibfnamefont{K.}~\bibnamefont{Kyutoku}},
  \bibinfo{author}{\bibfnamefont{M.}~\bibnamefont{Shibata}},
  \bibinfo{author}{\bibfnamefont{P.~R.} \bibnamefont{Brady}}, \bibnamefont{and}
  \bibinfo{author}{\bibfnamefont{J.~L.} \bibnamefont{Friedman}},
  \bibinfo{journal}{Phys. Rev.} \textbf{\bibinfo{volume}{D89}},
  \bibinfo{pages}{043009} (\bibinfo{year}{2014}), \eprint{1303.6298}.

\bibitem[{\citenamefont{Pannarale
  et~al.}(2015{\natexlab{b}})\citenamefont{Pannarale, Berti, Kyutoku, Lackey,
  and Shibata}}]{Pannarale:2015jka}
\bibinfo{author}{\bibfnamefont{F.}~\bibnamefont{Pannarale}},
  \bibinfo{author}{\bibfnamefont{E.}~\bibnamefont{Berti}},
  \bibinfo{author}{\bibfnamefont{K.}~\bibnamefont{Kyutoku}},
  \bibinfo{author}{\bibfnamefont{B.~D.} \bibnamefont{Lackey}},
  \bibnamefont{and} \bibinfo{author}{\bibfnamefont{M.}~\bibnamefont{Shibata}},
  \bibinfo{journal}{Phys. Rev.} \textbf{\bibinfo{volume}{D92}},
  \bibinfo{pages}{084050} (\bibinfo{year}{2015}{\natexlab{b}}),
  \eprint{1509.00512}.

\bibitem[{\citenamefont{{LIGO Scientific Collaboration}}(2018)}]{lalsuite}
\bibinfo{author}{\bibnamefont{{LIGO Scientific Collaboration}}},
  \emph{\bibinfo{title}{{LIGO} {A}lgorithm {L}ibrary - {LALS}uite}},
  \bibinfo{howpublished}{free software (GPL)} (\bibinfo{year}{2018}).

\bibitem[{\citenamefont{Hinderer}(2008)}]{Hinderer:2007mb}
\bibinfo{author}{\bibfnamefont{T.}~\bibnamefont{Hinderer}},
  \bibinfo{journal}{Astrophys. J.} \textbf{\bibinfo{volume}{677}},
  \bibinfo{pages}{1216} (\bibinfo{year}{2008}), \eprint{0711.2420}.

\bibitem[{\citenamefont{Santamaria et~al.}(2010)}]{Santamaria:2010yb}
\bibinfo{author}{\bibfnamefont{L.}~\bibnamefont{Santamaria}}
  \bibnamefont{et~al.}, \bibinfo{journal}{Phys. Rev.}
  \textbf{\bibinfo{volume}{D82}}, \bibinfo{pages}{064016}
  (\bibinfo{year}{2010}), \eprint{1005.3306}.

\bibitem[{\citenamefont{Husa et~al.}(2016)\citenamefont{Husa, Khan, Hannam,
  Pürrer, Ohme, Jiménez~Forteza, and Bohé}}]{Husa:2015iqa}
\bibinfo{author}{\bibfnamefont{S.}~\bibnamefont{Husa}},
  \bibinfo{author}{\bibfnamefont{S.}~\bibnamefont{Khan}},
  \bibinfo{author}{\bibfnamefont{M.}~\bibnamefont{Hannam}},
  \bibinfo{author}{\bibfnamefont{M.}~\bibnamefont{Pürrer}},
  \bibinfo{author}{\bibfnamefont{F.}~\bibnamefont{Ohme}},
  \bibinfo{author}{\bibfnamefont{X.}~\bibnamefont{Jiménez~Forteza}},
  \bibnamefont{and} \bibinfo{author}{\bibfnamefont{A.}~\bibnamefont{Bohé}},
  \bibinfo{journal}{Phys. Rev.} \textbf{\bibinfo{volume}{D93}},
  \bibinfo{pages}{044006} (\bibinfo{year}{2016}), \eprint{1508.07250}.

\bibitem[{\citenamefont{Khan et~al.}(2016)\citenamefont{Khan, Husa, Hannam,
  Ohme, Pürrer, Jiménez~Forteza, and Bohé}}]{Khan:2015jqa}
\bibinfo{author}{\bibfnamefont{S.}~\bibnamefont{Khan}},
  \bibinfo{author}{\bibfnamefont{S.}~\bibnamefont{Husa}},
  \bibinfo{author}{\bibfnamefont{M.}~\bibnamefont{Hannam}},
  \bibinfo{author}{\bibfnamefont{F.}~\bibnamefont{Ohme}},
  \bibinfo{author}{\bibfnamefont{M.}~\bibnamefont{Pürrer}},
  \bibinfo{author}{\bibfnamefont{X.}~\bibnamefont{Jiménez~Forteza}},
  \bibnamefont{and} \bibinfo{author}{\bibfnamefont{A.}~\bibnamefont{Bohé}},
  \bibinfo{journal}{Phys. Rev.} \textbf{\bibinfo{volume}{D93}},
  \bibinfo{pages}{044007} (\bibinfo{year}{2016}), \eprint{1508.07253}.

\bibitem[{\citenamefont{Varma and Ajith}(2017)}]{Varma:2016dnf}
\bibinfo{author}{\bibfnamefont{V.}~\bibnamefont{Varma}} \bibnamefont{and}
  \bibinfo{author}{\bibfnamefont{P.}~\bibnamefont{Ajith}},
  \bibinfo{journal}{Phys. Rev.} \textbf{\bibinfo{volume}{D96}},
  \bibinfo{pages}{124024} (\bibinfo{year}{2017}), \eprint{1612.05608}.

\bibitem[{\citenamefont{Kalaghatgi et~al.}(2019)\citenamefont{Kalaghatgi,
  Hannam, and Raymond}}]{Kalaghatgi:2019log}
\bibinfo{author}{\bibfnamefont{C.}~\bibnamefont{Kalaghatgi}},
  \bibinfo{author}{\bibfnamefont{M.}~\bibnamefont{Hannam}}, \bibnamefont{and}
  \bibinfo{author}{\bibfnamefont{V.}~\bibnamefont{Raymond}}
  (\bibinfo{year}{2019}), \eprint{1909.10010}.

\bibitem[{\citenamefont{García-Quirós
  et~al.}(2020)\citenamefont{García-Quirós, Colleoni, Husa, Estellés,
  Pratten, Ramos-Buades, Mateu-Lucena, and Jaume}}]{Garcia-Quiros:2020qpx}
\bibinfo{author}{\bibfnamefont{C.}~\bibnamefont{García-Quirós}},
  \bibinfo{author}{\bibfnamefont{M.}~\bibnamefont{Colleoni}},
  \bibinfo{author}{\bibfnamefont{S.}~\bibnamefont{Husa}},
  \bibinfo{author}{\bibfnamefont{H.}~\bibnamefont{Estellés}},
  \bibinfo{author}{\bibfnamefont{G.}~\bibnamefont{Pratten}},
  \bibinfo{author}{\bibfnamefont{A.}~\bibnamefont{Ramos-Buades}},
  \bibinfo{author}{\bibfnamefont{M.}~\bibnamefont{Mateu-Lucena}},
  \bibnamefont{and} \bibinfo{author}{\bibfnamefont{R.}~\bibnamefont{Jaume}}
  (\bibinfo{year}{2020}), \eprint{2001.10914}.

\bibitem[{\citenamefont{Read et~al.}(2009)\citenamefont{Read, Markakis,
  Shibata, Ury\ifmmode~\bar{u}\else \={u}\fi{}, Creighton, and
  Friedman}}]{PhysRevD.79.124033}
\bibinfo{author}{\bibfnamefont{J.~S.} \bibnamefont{Read}},
  \bibinfo{author}{\bibfnamefont{C.}~\bibnamefont{Markakis}},
  \bibinfo{author}{\bibfnamefont{M.}~\bibnamefont{Shibata}},
  \bibinfo{author}{\bibfnamefont{K.~b.~o.}
  \bibnamefont{Ury\ifmmode~\bar{u}\else \={u}\fi{}}},
  \bibinfo{author}{\bibfnamefont{J.~D.~E.} \bibnamefont{Creighton}},
  \bibnamefont{and} \bibinfo{author}{\bibfnamefont{J.~L.}
  \bibnamefont{Friedman}}, \bibinfo{journal}{Phys. Rev. D}
  \textbf{\bibinfo{volume}{79}}, \bibinfo{pages}{124033}
  (\bibinfo{year}{2009}),
  \urlprefix\url{https://link.aps.org/doi/10.1103/PhysRevD.79.124033}.

\bibitem[{\citenamefont{Tolman}(1934)}]{Tolman169}
\bibinfo{author}{\bibfnamefont{R.~C.} \bibnamefont{Tolman}},
  \bibinfo{journal}{Proceedings of the National Academy of Sciences}
  \textbf{\bibinfo{volume}{20}}, \bibinfo{pages}{169} (\bibinfo{year}{1934}),
  ISSN \bibinfo{issn}{0027-8424},
  \eprint{https://www.pnas.org/content/20/3/169.full.pdf},
  \urlprefix\url{https://www.pnas.org/content/20/3/169}.

\bibitem[{\citenamefont{Tolman}(1939)}]{PhysRev.55.364}
\bibinfo{author}{\bibfnamefont{R.~C.} \bibnamefont{Tolman}},
  \bibinfo{journal}{Phys. Rev.} \textbf{\bibinfo{volume}{55}},
  \bibinfo{pages}{364} (\bibinfo{year}{1939}),
  \urlprefix\url{https://link.aps.org/doi/10.1103/PhysRev.55.364}.

\bibitem[{\citenamefont{Oppenheimer and Volkoff}(1939)}]{PhysRev.55.374}
\bibinfo{author}{\bibfnamefont{J.~R.} \bibnamefont{Oppenheimer}}
  \bibnamefont{and} \bibinfo{author}{\bibfnamefont{G.~M.}
  \bibnamefont{Volkoff}}, \bibinfo{journal}{Phys. Rev.}
  \textbf{\bibinfo{volume}{55}}, \bibinfo{pages}{374} (\bibinfo{year}{1939}),
  \urlprefix\url{https://link.aps.org/doi/10.1103/PhysRev.55.374}.

\bibitem[{\citenamefont{Lackey and Wade}(2015)}]{Lackey:2014fwa}
\bibinfo{author}{\bibfnamefont{B.~D.} \bibnamefont{Lackey}} \bibnamefont{and}
  \bibinfo{author}{\bibfnamefont{L.}~\bibnamefont{Wade}},
  \bibinfo{journal}{Phys. Rev.} \textbf{\bibinfo{volume}{D91}},
  \bibinfo{pages}{043002} (\bibinfo{year}{2015}), \eprint{1410.8866}.

\bibitem[{\citenamefont{Pannarale et~al.}(2013)\citenamefont{Pannarale, Berti,
  Kyutoku, and Shibata}}]{Pannarale:2013uoa}
\bibinfo{author}{\bibfnamefont{F.}~\bibnamefont{Pannarale}},
  \bibinfo{author}{\bibfnamefont{E.}~\bibnamefont{Berti}},
  \bibinfo{author}{\bibfnamefont{K.}~\bibnamefont{Kyutoku}}, \bibnamefont{and}
  \bibinfo{author}{\bibfnamefont{M.}~\bibnamefont{Shibata}},
  \bibinfo{journal}{Phys. Rev.} \textbf{\bibinfo{volume}{D88}},
  \bibinfo{pages}{084011} (\bibinfo{year}{2013}), \eprint{1307.5111}.

\bibitem[{\citenamefont{Chakravarti et~al.}(2019)}]{Chakravarti:2018uyi}
\bibinfo{author}{\bibfnamefont{K.}~\bibnamefont{Chakravarti}}
  \bibnamefont{et~al.}, \bibinfo{journal}{Phys. Rev.}
  \textbf{\bibinfo{volume}{D99}}, \bibinfo{pages}{024049}
  (\bibinfo{year}{2019}), \eprint{1809.04349}.

\bibitem[{ali()}]{aligozerodethp}
\bibinfo{howpublished}{\url{https://dcc.ligo.org/LIGO-T0900288/public}},
  \urlprefix\url{https://dcc.ligo.org/LIGO-T0900288/public}.

\bibitem[{\citenamefont{Taracchini et~al.}(2012)\citenamefont{Taracchini, Pan,
  Buonanno, Barausse, Boyle, Chu, Lovelace, Pfeiffer, and
  Scheel}}]{PhysRevD.86.024011}
\bibinfo{author}{\bibfnamefont{A.}~\bibnamefont{Taracchini}},
  \bibinfo{author}{\bibfnamefont{Y.}~\bibnamefont{Pan}},
  \bibinfo{author}{\bibfnamefont{A.}~\bibnamefont{Buonanno}},
  \bibinfo{author}{\bibfnamefont{E.}~\bibnamefont{Barausse}},
  \bibinfo{author}{\bibfnamefont{M.}~\bibnamefont{Boyle}},
  \bibinfo{author}{\bibfnamefont{T.}~\bibnamefont{Chu}},
  \bibinfo{author}{\bibfnamefont{G.}~\bibnamefont{Lovelace}},
  \bibinfo{author}{\bibfnamefont{H.~P.} \bibnamefont{Pfeiffer}},
  \bibnamefont{and} \bibinfo{author}{\bibfnamefont{M.~A.}
  \bibnamefont{Scheel}}, \bibinfo{journal}{Phys. Rev. D}
  \textbf{\bibinfo{volume}{86}}, \bibinfo{pages}{024011}
  (\bibinfo{year}{2012}),
  \urlprefix\url{https://link.aps.org/doi/10.1103/PhysRevD.86.024011}.

\bibitem[{\citenamefont{Taracchini et~al.}(2014)\citenamefont{Taracchini,
  Buonanno, Pan, Hinderer, Boyle, Hemberger, Kidder, Lovelace, Mrou\'e,
  Pfeiffer et~al.}}]{PhysRevD.89.061502}
\bibinfo{author}{\bibfnamefont{A.}~\bibnamefont{Taracchini}},
  \bibinfo{author}{\bibfnamefont{A.}~\bibnamefont{Buonanno}},
  \bibinfo{author}{\bibfnamefont{Y.}~\bibnamefont{Pan}},
  \bibinfo{author}{\bibfnamefont{T.}~\bibnamefont{Hinderer}},
  \bibinfo{author}{\bibfnamefont{M.}~\bibnamefont{Boyle}},
  \bibinfo{author}{\bibfnamefont{D.~A.} \bibnamefont{Hemberger}},
  \bibinfo{author}{\bibfnamefont{L.~E.} \bibnamefont{Kidder}},
  \bibinfo{author}{\bibfnamefont{G.}~\bibnamefont{Lovelace}},
  \bibinfo{author}{\bibfnamefont{A.~H.} \bibnamefont{Mrou\'e}},
  \bibinfo{author}{\bibfnamefont{H.~P.} \bibnamefont{Pfeiffer}},
  \bibnamefont{et~al.}, \bibinfo{journal}{Phys. Rev. D}
  \textbf{\bibinfo{volume}{89}}, \bibinfo{pages}{061502}
  (\bibinfo{year}{2014}),
  \urlprefix\url{https://link.aps.org/doi/10.1103/PhysRevD.89.061502}.

\bibitem[{\citenamefont{Ohme et~al.}(2011)\citenamefont{Ohme, Hannam, and
  Husa}}]{Ohme:2011zm}
\bibinfo{author}{\bibfnamefont{F.}~\bibnamefont{Ohme}},
  \bibinfo{author}{\bibfnamefont{M.}~\bibnamefont{Hannam}}, \bibnamefont{and}
  \bibinfo{author}{\bibfnamefont{S.}~\bibnamefont{Husa}},
  \bibinfo{journal}{Phys. Rev.} \textbf{\bibinfo{volume}{D84}},
  \bibinfo{pages}{064029} (\bibinfo{year}{2011}), \eprint{1107.0996}.

\bibitem[{\citenamefont{{Boh{\'e}} et~al.}(2017)\citenamefont{{Boh{\'e}},
  {Shao}, {Taracchini}, {Buonanno}, {Babak}, {Harry}, {Hinder}, {Ossokine},
  {P{\"u}rrer}, {Raymond} et~al.}}]{2017PhRvD..95d4028B}
\bibinfo{author}{\bibfnamefont{A.}~\bibnamefont{{Boh{\'e}}}},
  \bibinfo{author}{\bibfnamefont{L.}~\bibnamefont{{Shao}}},
  \bibinfo{author}{\bibfnamefont{A.}~\bibnamefont{{Taracchini}}},
  \bibinfo{author}{\bibfnamefont{A.}~\bibnamefont{{Buonanno}}},
  \bibinfo{author}{\bibfnamefont{S.}~\bibnamefont{{Babak}}},
  \bibinfo{author}{\bibfnamefont{I.~W.} \bibnamefont{{Harry}}},
  \bibinfo{author}{\bibfnamefont{I.}~\bibnamefont{{Hinder}}},
  \bibinfo{author}{\bibfnamefont{S.}~\bibnamefont{{Ossokine}}},
  \bibinfo{author}{\bibfnamefont{M.}~\bibnamefont{{P{\"u}rrer}}},
  \bibinfo{author}{\bibfnamefont{V.}~\bibnamefont{{Raymond}}},
  \bibnamefont{et~al.}, \bibinfo{journal}{\prd} \textbf{\bibinfo{volume}{95}},
  \bibinfo{eid}{044028} (\bibinfo{year}{2017}), \eprint{1611.03703}.

\bibitem[{\citenamefont{Hotokezaka et~al.}(2016)\citenamefont{Hotokezaka,
  Kyutoku, Sekiguchi, and Shibata}}]{Hotokezaka:2016bzh}
\bibinfo{author}{\bibfnamefont{K.}~\bibnamefont{Hotokezaka}},
  \bibinfo{author}{\bibfnamefont{K.}~\bibnamefont{Kyutoku}},
  \bibinfo{author}{\bibfnamefont{Y.-i.} \bibnamefont{Sekiguchi}},
  \bibnamefont{and} \bibinfo{author}{\bibfnamefont{M.}~\bibnamefont{Shibata}},
  \bibinfo{journal}{Phys. Rev.} \textbf{\bibinfo{volume}{D93}},
  \bibinfo{pages}{064082} (\bibinfo{year}{2016}), \eprint{1603.01286}.

\bibitem[{\citenamefont{Baird et~al.}(2013)\citenamefont{Baird, Fairhurst,
  Hannam, and Murphy}}]{Baird:2012cu}
\bibinfo{author}{\bibfnamefont{E.}~\bibnamefont{Baird}},
  \bibinfo{author}{\bibfnamefont{S.}~\bibnamefont{Fairhurst}},
  \bibinfo{author}{\bibfnamefont{M.}~\bibnamefont{Hannam}}, \bibnamefont{and}
  \bibinfo{author}{\bibfnamefont{P.}~\bibnamefont{Murphy}},
  \bibinfo{journal}{Phys. Rev.} \textbf{\bibinfo{volume}{D87}},
  \bibinfo{pages}{024035} (\bibinfo{year}{2013}), \eprint{1211.0546}.

\bibitem[{\citenamefont{Punturo et~al.}(2010)}]{Punturo:2010zza}
\bibinfo{author}{\bibfnamefont{M.}~\bibnamefont{Punturo}} \bibnamefont{et~al.},
  \bibinfo{journal}{Class. Quant. Grav.} \textbf{\bibinfo{volume}{27}},
  \bibinfo{pages}{084007} (\bibinfo{year}{2010}).

\bibitem[{\citenamefont{Hild et~al.}(2011)}]{Hild:2010id}
\bibinfo{author}{\bibfnamefont{S.}~\bibnamefont{Hild}} \bibnamefont{et~al.},
  \bibinfo{journal}{Class. Quant. Grav.} \textbf{\bibinfo{volume}{28}},
  \bibinfo{pages}{094013} (\bibinfo{year}{2011}), \eprint{1012.0908}.

\bibitem[{\citenamefont{Abbott et~al.}(2017)}]{Evans:2016mbw}
\bibinfo{author}{\bibfnamefont{B.~P.} \bibnamefont{Abbott}}
  \bibnamefont{et~al.} (\bibinfo{collaboration}{LIGO Scientific}),
  \bibinfo{journal}{Class. Quant. Grav.} \textbf{\bibinfo{volume}{34}},
  \bibinfo{pages}{044001} (\bibinfo{year}{2017}), \eprint{1607.08697}.

\bibitem[{\citenamefont{Mapelli and Giacobbo}(2018)}]{10.1093/mnras/sty1613}
\bibinfo{author}{\bibfnamefont{M.}~\bibnamefont{Mapelli}} \bibnamefont{and}
  \bibinfo{author}{\bibfnamefont{N.}~\bibnamefont{Giacobbo}},
  \bibinfo{journal}{Monthly Notices of the Royal Astronomical Society}
  \textbf{\bibinfo{volume}{479}}, \bibinfo{pages}{4391} (\bibinfo{year}{2018}),
  ISSN \bibinfo{issn}{0035-8711},
  \eprint{https://academic.oup.com/mnras/article-pdf/479/4/4391/25180521/sty1613.pdf},
  \urlprefix\url{https://doi.org/10.1093/mnras/sty1613}.

\bibitem[{\citenamefont{Apostolatos et~al.}(1994)\citenamefont{Apostolatos,
  Cutler, Sussman, and Thorne}}]{Apostolatos:1994mx}
\bibinfo{author}{\bibfnamefont{T.~A.} \bibnamefont{Apostolatos}},
  \bibinfo{author}{\bibfnamefont{C.}~\bibnamefont{Cutler}},
  \bibinfo{author}{\bibfnamefont{G.~J.} \bibnamefont{Sussman}},
  \bibnamefont{and} \bibinfo{author}{\bibfnamefont{K.~S.}
  \bibnamefont{Thorne}}, \bibinfo{journal}{Phys. Rev.}
  \textbf{\bibinfo{volume}{D49}}, \bibinfo{pages}{6274} (\bibinfo{year}{1994}).

\bibitem[{\citenamefont{Apostolatos}(1995)}]{Apostolatos:1995pj}
\bibinfo{author}{\bibfnamefont{T.~A.} \bibnamefont{Apostolatos}},
  \bibinfo{journal}{Phys. Rev.} \textbf{\bibinfo{volume}{D52}},
  \bibinfo{pages}{605} (\bibinfo{year}{1995}).

\bibitem[{\citenamefont{Kidder et~al.}(1993)\citenamefont{Kidder, Will, and
  Wiseman}}]{Kidder:1993zz}
\bibinfo{author}{\bibfnamefont{L.~E.} \bibnamefont{Kidder}},
  \bibinfo{author}{\bibfnamefont{C.~M.} \bibnamefont{Will}}, \bibnamefont{and}
  \bibinfo{author}{\bibfnamefont{A.~G.} \bibnamefont{Wiseman}},
  \bibinfo{journal}{Phys. Rev.} \textbf{\bibinfo{volume}{D47}},
  \bibinfo{pages}{3281} (\bibinfo{year}{1993}).

\bibitem[{\citenamefont{Hunter}(2007)}]{Hunter:2007}
\bibinfo{author}{\bibfnamefont{J.~D.} \bibnamefont{Hunter}},
  \bibinfo{journal}{Computing in Science \& Engineering}
  \textbf{\bibinfo{volume}{9}}, \bibinfo{pages}{90} (\bibinfo{year}{2007}).

\bibitem[{\citenamefont{{van der Walt} et~al.}(2011)\citenamefont{{van der
  Walt}, {Colbert}, and {Varoquaux}}}]{5725236}
\bibinfo{author}{\bibfnamefont{S.}~\bibnamefont{{van der Walt}}},
  \bibinfo{author}{\bibfnamefont{S.~C.} \bibnamefont{{Colbert}}},
  \bibnamefont{and}
  \bibinfo{author}{\bibfnamefont{G.}~\bibnamefont{{Varoquaux}}},
  \bibinfo{journal}{Computing in Science Engineering}
  \textbf{\bibinfo{volume}{13}}, \bibinfo{pages}{22} (\bibinfo{year}{2011}),
  ISSN \bibinfo{issn}{1558-366X}.

\bibitem[{\citenamefont{Nitz et~al.}(2020)\citenamefont{Nitz, Harry, Brown,
  Biwer, Willis, Canton, Capano, Pekowsky, Dent, Williamson
  et~al.}}]{alex_nitz_2020_3630601}
\bibinfo{author}{\bibfnamefont{A.}~\bibnamefont{Nitz}},
  \bibinfo{author}{\bibfnamefont{I.}~\bibnamefont{Harry}},
  \bibinfo{author}{\bibfnamefont{D.}~\bibnamefont{Brown}},
  \bibinfo{author}{\bibfnamefont{C.~M.} \bibnamefont{Biwer}},
  \bibinfo{author}{\bibfnamefont{J.}~\bibnamefont{Willis}},
  \bibinfo{author}{\bibfnamefont{T.~D.} \bibnamefont{Canton}},
  \bibinfo{author}{\bibfnamefont{C.}~\bibnamefont{Capano}},
  \bibinfo{author}{\bibfnamefont{L.}~\bibnamefont{Pekowsky}},
  \bibinfo{author}{\bibfnamefont{T.}~\bibnamefont{Dent}},
  \bibinfo{author}{\bibfnamefont{A.~R.} \bibnamefont{Williamson}},
  \bibnamefont{et~al.}, \emph{\bibinfo{title}{gwastro/pycbc: Pycbc release
  v1.15.4}} (\bibinfo{year}{2020}),
  \urlprefix\url{https://doi.org/10.5281/zenodo.3630601}.

\bibitem[{\citenamefont{{Oliphant}}(2007)}]{4160250}
\bibinfo{author}{\bibfnamefont{T.~E.} \bibnamefont{{Oliphant}}},
  \bibinfo{journal}{Computing in Science Engineering}
  \textbf{\bibinfo{volume}{9}}, \bibinfo{pages}{10} (\bibinfo{year}{2007}),
  ISSN \bibinfo{issn}{1558-366X}.

\bibitem[{\citenamefont{Foucart}(2012)}]{Foucart:2012nc}
\bibinfo{author}{\bibfnamefont{F.}~\bibnamefont{Foucart}},
  \bibinfo{journal}{Phys. Rev.} \textbf{\bibinfo{volume}{D86}},
  \bibinfo{pages}{124007} (\bibinfo{year}{2012}), \eprint{1207.6304}.

\bibitem[{\citenamefont{Shibata and Taniguchi}(2008)}]{Shibata:2007zm}
\bibinfo{author}{\bibfnamefont{M.}~\bibnamefont{Shibata}} \bibnamefont{and}
  \bibinfo{author}{\bibfnamefont{K.}~\bibnamefont{Taniguchi}},
  \bibinfo{journal}{Phys. Rev.} \textbf{\bibinfo{volume}{D77}},
  \bibinfo{pages}{084015} (\bibinfo{year}{2008}), \eprint{0711.1410}.

\bibitem[{\citenamefont{{Bardeen} et~al.}(1972)\citenamefont{{Bardeen},
  {Press}, and {Teukolsky}}}]{1972ApJ...178..347B}
\bibinfo{author}{\bibfnamefont{J.~M.} \bibnamefont{{Bardeen}}},
  \bibinfo{author}{\bibfnamefont{W.~H.} \bibnamefont{{Press}}},
  \bibnamefont{and} \bibinfo{author}{\bibfnamefont{S.~A.}
  \bibnamefont{{Teukolsky}}}, \bibinfo{journal}{\apj}
  \textbf{\bibinfo{volume}{178}}, \bibinfo{pages}{347} (\bibinfo{year}{1972}).

\bibitem[{\citenamefont{Foucart et~al.}(2018)\citenamefont{Foucart, Hinderer,
  and Nissanke}}]{Foucart:2018rjc}
\bibinfo{author}{\bibfnamefont{F.}~\bibnamefont{Foucart}},
  \bibinfo{author}{\bibfnamefont{T.}~\bibnamefont{Hinderer}}, \bibnamefont{and}
  \bibinfo{author}{\bibfnamefont{S.}~\bibnamefont{Nissanke}},
  \bibinfo{journal}{Phys. Rev.} \textbf{\bibinfo{volume}{D98}},
  \bibinfo{pages}{081501} (\bibinfo{year}{2018}), \eprint{1807.00011}.

\bibitem[{\citenamefont{Zappa et~al.}(2019)\citenamefont{Zappa, Bernuzzi,
  Pannarale, Mapelli, and Giacobbo}}]{Zappa:2019ntl}
\bibinfo{author}{\bibfnamefont{F.}~\bibnamefont{Zappa}},
  \bibinfo{author}{\bibfnamefont{S.}~\bibnamefont{Bernuzzi}},
  \bibinfo{author}{\bibfnamefont{F.}~\bibnamefont{Pannarale}},
  \bibinfo{author}{\bibfnamefont{M.}~\bibnamefont{Mapelli}}, \bibnamefont{and}
  \bibinfo{author}{\bibfnamefont{N.}~\bibnamefont{Giacobbo}},
  \bibinfo{journal}{Phys. Rev. Lett.} \textbf{\bibinfo{volume}{123}},
  \bibinfo{pages}{041102} (\bibinfo{year}{2019}), \eprint{1903.11622}.

\bibitem[{\citenamefont{Jiménez-Forteza
  et~al.}(2017)\citenamefont{Jiménez-Forteza, Keitel, Husa, Hannam, Khan, and
  Pürrer}}]{Jimenez-Forteza:2016oae}
\bibinfo{author}{\bibfnamefont{X.}~\bibnamefont{Jiménez-Forteza}},
  \bibinfo{author}{\bibfnamefont{D.}~\bibnamefont{Keitel}},
  \bibinfo{author}{\bibfnamefont{S.}~\bibnamefont{Husa}},
  \bibinfo{author}{\bibfnamefont{M.}~\bibnamefont{Hannam}},
  \bibinfo{author}{\bibfnamefont{S.}~\bibnamefont{Khan}}, \bibnamefont{and}
  \bibinfo{author}{\bibfnamefont{M.}~\bibnamefont{Pürrer}},
  \bibinfo{journal}{Phys. Rev.} \textbf{\bibinfo{volume}{D95}},
  \bibinfo{pages}{064024} (\bibinfo{year}{2017}), \eprint{1611.00332}.

\bibitem[{\citenamefont{London and Fauchon-Jones}(2018)}]{London:2018nxs}
\bibinfo{author}{\bibfnamefont{L.}~\bibnamefont{London}} \bibnamefont{and}
  \bibinfo{author}{\bibfnamefont{E.}~\bibnamefont{Fauchon-Jones}}
  (\bibinfo{year}{2018}), \eprint{1810.03550}.

\bibitem[{\citenamefont{Buonanno et~al.}(2003)\citenamefont{Buonanno, Chen, and
  Vallisneri}}]{Buonanno:2002fy}
\bibinfo{author}{\bibfnamefont{A.}~\bibnamefont{Buonanno}},
  \bibinfo{author}{\bibfnamefont{Y.-b.} \bibnamefont{Chen}}, \bibnamefont{and}
  \bibinfo{author}{\bibfnamefont{M.}~\bibnamefont{Vallisneri}},
  \bibinfo{journal}{Phys. Rev.} \textbf{\bibinfo{volume}{D67}},
  \bibinfo{pages}{104025} (\bibinfo{year}{2003}), \bibinfo{note}{[Erratum:
  Phys. Rev.D74,029904(2006)]}, \eprint{gr-qc/0211087}.

\bibitem[{\citenamefont{Yagi and Yunes}(2017)}]{Yagi:2016bkt}
\bibinfo{author}{\bibfnamefont{K.}~\bibnamefont{Yagi}} \bibnamefont{and}
  \bibinfo{author}{\bibfnamefont{N.}~\bibnamefont{Yunes}},
  \bibinfo{journal}{Phys. Rept.} \textbf{\bibinfo{volume}{681}},
  \bibinfo{pages}{1} (\bibinfo{year}{2017}), \eprint{1608.02582}.

\bibitem[{\citenamefont{Matas et~al.}(2020)\citenamefont{Matas, Buonanno,
  Dietrich, and Hinderer}}]{SEOBNRNSBH}
\bibinfo{author}{\bibfnamefont{A.}~\bibnamefont{Matas}},
  \bibinfo{author}{\bibfnamefont{A.}~\bibnamefont{Buonanno}},
  \bibinfo{author}{\bibfnamefont{T.}~\bibnamefont{Dietrich}}, \bibnamefont{and}
  \bibinfo{author}{\bibfnamefont{T.}~\bibnamefont{Hinderer}}
  (\bibinfo{year}{2020}), \bibinfo{note}{in preparation}.

\end{thebibliography}
\end{document}